\documentclass[aps,preprint,superscriptaddress,longbibliography]{revtex4-1}

\usepackage{graphicx}
\usepackage{color}
\usepackage{amsmath, amssymb}
\usepackage{latexsym}

\newcommand{\ie}{\textit{i.e.}}
\newcommand{\eg}{\textit{e.g.}}
\newcommand{\Ssps}{${\it\sigma_{SPS}}$}
\newcommand{\Spps}{${\it\sigma_{PPS}}$}
\newcommand{\G}{${\it\Gamma}$}
\newcommand{\Gsps}{${\it\Gamma_{SPS}}$}
\newcommand{\Gpps}{${\it\Gamma_{PPS}}$}
\newcommand{\Gcpps}{${\it\Gamma_{CPPS}}$}
\newcommand{\Gqpps}{${\it\Gamma_{QPPS}}$}
\newcommand{\Psps}{${\it P_{SPS}}$}

\newcommand{\Pcpps}{${\it P_{CPPS}}$}

\begin{document}


\title{Formation of Quantum Phase Slip Pairs in Superconducting Nanowires}

\author{A. Belkin}
\affiliation{Department of Physics, University of Illinois at Urbana-Champaign, Urbana IL 61801}

\author{M. Belkin}
\affiliation{Department of Physics, University of Illinois at Urbana-Champaign, Urbana IL 61801}

\author{V. Vakaryuk}
\affiliation{Department of Physics and Astronomy, Johns Hopkins University, Baltimore, MD 21218}
\affiliation{American Physical Society, 1 Research Road, Ridge, NY 11961}

\author{S. Khlebnikov}
\affiliation{Department of Physics and Astronomy, Purdue University, West Lafayette, IN 47907}

\author{A. Bezryadin} 
\email[]{bezryadi@illinois.edu}
\affiliation{Department of Physics, University of Illinois at Urbana-Champaign, Urbana IL 61801}

\date{\today}

\begin{abstract}

Macroscopic quantum tunneling (MQT) is a fundamental phenomenon of quantum mechanics related to the actively debated topic of quantum-to-classical transition. The ability to realize MQT affects implementation of qubit-based quantum computing schemes and their protection against decoherence. Decoherence in qubits can be reduced by means of topological protection, {\it e.g.} by exploiting various parity effects. In particular, paired phase slips can provide such protection for superconducting qubits. Here, we report on the direct observation of quantum paired phase slips in thin-wire superconducting loops. We show that in addition to conventional single phase slips that change superconducting order parameter phase by $2\pi$, there are quantum transitions changing the phase by $4\pi$. Quantum paired phase slips represent a synchronized occurrence of two macroscopic quantum tunneling events, \ie\ cotunneling. We demonstrate the existence of a remarkable regime in which paired phase slips are exponentially more probable than single ones.

\end{abstract}

\maketitle

Decoherence and quantum noise are primary roadblocks for large-scale implementation of quantum computing that in many cases relies on the phenomenon of macroscopic quantum tunneling~\cite{Leggett:1980, Lee:2011, Clarke:1988, Jackel:1981,Wernsdorfer:1997}. Decoherence is inherently related to the collapse of macroscopic wave function caused by various environment induced interactions. In quantum computers detrimental effects of decoherence can be eliminated by means of quantum error correction as originally proposed by Shor~\cite{Shor:1995}. An alternative approach was suggested by Kitaev~\cite{Kitaev:2003}, who argued that the use of topologically ordered quantum systems can eliminate the burden of quantum error correction. In such systems qubits are implemented by topologically distinct states connected by a global operation (such as braiding in case of anyons) that cannot be mixed by a local perturbation.

Less exotic schemes of topological protection based on various parity effects in superconducting circuits have also been presented~\cite{Gottesman:2001,Dou:2002,Gladchenko:2009, Bell:2014}. The crucial component of these proposals is a device that discriminates between parity conserving and parity violating transitions. Ideally, such a device fully suppresses the latter, thus creating well separated parity-based sectors in the Hilbert space, which are used as a ``grid'' for qubit operations.

Here, we focus on a  multiply-connected device formed by a superconducting loop containing homogeneous superconducting nanowires. Flux states of the loop are described by the winding number (vorticity) of the superconducting phase on a path encircling the loop. Transitions between different winding number states occur through phase slips taking place in the nanowires. Realization of a parity-protected qubit requires suppression of $2\pi$ phase slips, which change the winding number by one, and a significant amplitude of $4\pi$ phase slips, which are parity conserving~\cite{Dou:2012,Guinea:1986, Korshunov:1987} events. We term them single phase slips (SPS) and paired phase slips (PPS) respectively. 

Historically, the phase slips were proposed by Little~\cite{Little:1967} as thermally activated topological transitions in the space of the winding numbers of the superconducting condensate in a one-dimensional wire. At low temperatures, such events should proceed through MQT. The unique feature of an MQT in thin wires (as opposed to tunnel junctions) is that normal electrons at the core of the phase slip create an additional dissipative component at the location of the tunneling process. One may question if MQT is still possible in these circumstances. So far, demonstrations of quantum phase slips in homogeneous wires either relied on indirect evidences, such as overheating of a wire by a phase slip~\cite{Sahu:2009,Li:2011,Aref:2012,Bezryadin:2012}, or were related to the collective effects of many phase slips~\cite{Giordano:1988,Bezryadin:2000,Lau:2001,Arutyunov:2008}. The important advancement has been recently made, showing delocalization of a wave-function due to a coherent tunneling of multiple phase slips in strongly disordered InO$_{\mathrm{x}}$~\cite{Astafiev:2012} and NbN~\cite{Peltonen:2013} nanowires.

In the present work, we study MQT in thin-wire superconducting loops by means of a microwave measurement technique that is capable of detecting individual phase slips~\cite{Belkin:2011}. We observe individual phase slips as isolated macroscopic tunneling events. We demonstrate the presence of quantum-paired phase slips (QPPS), which change the loop vorticity by 2, and hence, preserve the parity of the winding quantum number. Such transitions correspond to simultaneous tunneling (``cotunneling'') of two fluxoids {\it in} or {\it out} of the loop. They can be thought of as result of quantum synchronization, or pairing, of distinct macroscopic events that are governed by the minimization of action rather than the minimization of energy. More importantly, we discover that QPPS are exponentially more likely than SPS if the bias is lower than a certain critical value. In this regime our thin-wire superconducting loop acts as a parity conserving element.  We argue that the observed dominance of QPPS is caused by their weaker coupling to gapless environmental modes and their lower effective mass, compared to SPS. We note that although phase slips including quantum ones have been studied by many research groups, phase slip pairing observed here has not been previously reported to the best of our knowledge.

From a theoretical point of view the study of quantum-paired phase slips is motivated in part by their relevance to parity protected qubits for future generations of quantum computers ~\cite{Dou:2012}. In such hypothetical devices  quantum information could be encoded in the parity of the winding number of a superconducting loop. If a device in which all phase slips are paired could be realized, its parity would not be affected by quantum fluctuations of the vorticity, because single phase slips would be suppressed and paired phase slips cannot change the parity. While in the present work we do not observe any qubit effects we nevertheless identify a regime in which paired phase slips are much more frequent than single phase slips. Additionally, we discuss theoretically a physical mechanism behind the dominance of paired phase slips.

\begin{figure}[htbp]
\centering
	\includegraphics[width=86mm]{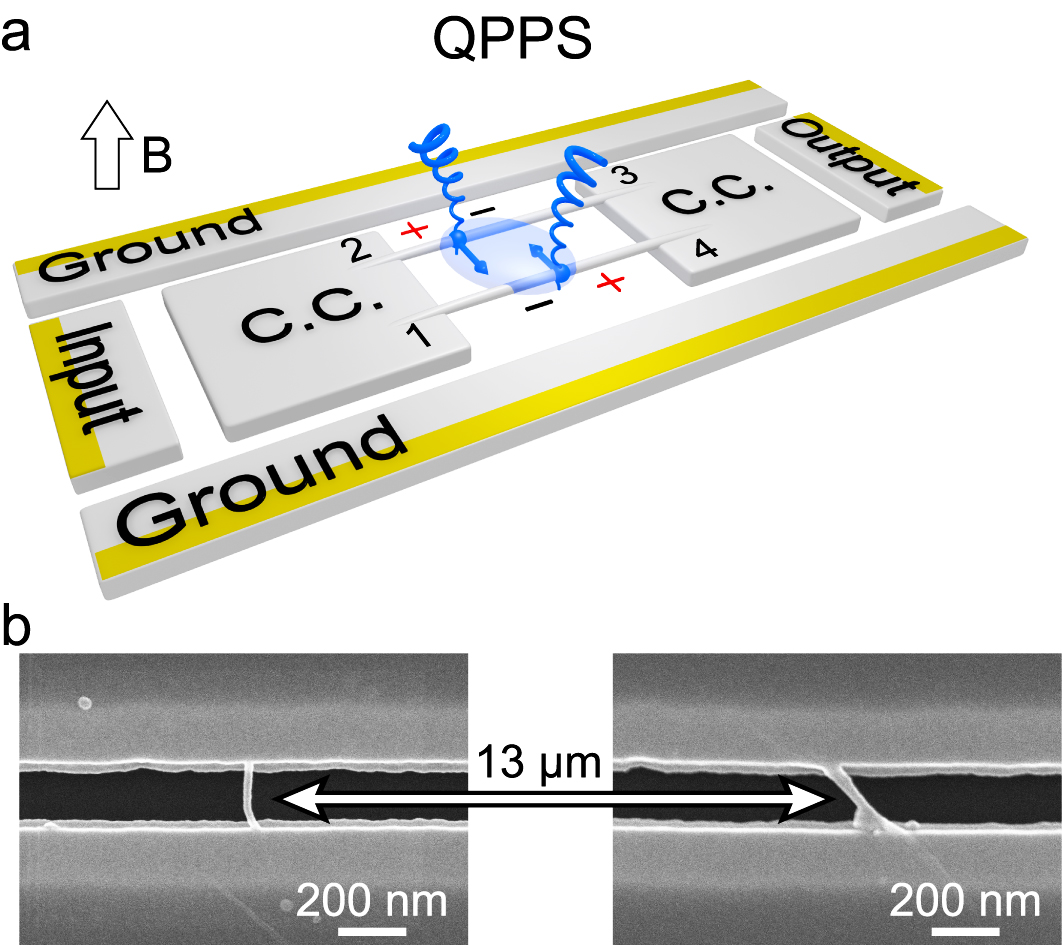}
	\caption{Schematic representation of the samples (not to scale). (a) Superconducting coplanar resonator has two nanowires in the middle of the center conductor, which are suspended over a trench in the substrate (see SM-1). MoGe film is shown in gray, regions where it is removed -- in white. The superconducting loop is threaded by an external magnetic field, $B$. Yellow stripes depict gold coating. QPPS is shown as a pair of fluxoids simultaneously crossing both nanowires. Each fluxoid changes the phase accumulated on the loop by $2\pi$ and generates voltage on each wire $V\!=\!\hbar(d\phi_{wire}/dt)/2e$. Plus and minus signs indicate the quadrupole charge distribution during QPPS. (b) Scanning electron microscope images of the nanowires in sample A. The spacing between nanowires is 13~$\mu m$. } 
	\label{fig1}
\end{figure}

The geometry of our samples is illustrated in Fig.~\ref{fig1} (see also Ref.~\cite{Belkin:2011} for additional technical details on the sample design and the measurement setup). A pair of superconducting nanowires fabricated by molecular templating~\cite{Bezryadin:2000,Bezryadin:2012} is integrated into the middle of the center conductor (c.c.) of a superconducting coplanar waveguide Fabry-P\'{e}rot resonator (see SM-1). The wires together with the two halves of c.c.~strip form a closed superconducting loop. An external magnetic field, $B$, applied perpendicular to the resonator's surface induces Meissner screening currents in the c.c.~strips. These currents produce a field-dependent phase difference of the superconducting order parameter between points of contact of the wires and the center conductor, which we denote $\phi_{12}(B)$ and $\phi_{34}(B)$ (see Fig.~\ref{fig1}). Single-valuedness requirement for the order parameter leads to the following quantization condition: $\phi_{12}\!+\!\phi_{23}\!+\!\phi_{34}\!+\!\phi_{41}\!=\!2\pi n$. Here, $\phi_{23}$ and $\phi_{41}$ are phases accumulated along the wires in passing from point 2 to 3, and 4 to 1 respectively (see Fig.~\ref{fig1}); $n$ is vorticity, \ie\ the number of fluxoids trapped in the loop. Since the c.c.~strips and both wires are almost identical the respective phase accumulations are assumed equal, \ie\ $\phi_{12}\!=\!\phi_{34}\!=\!\phi_{cc}$, $\phi_{23}\!=\!\phi_{41}=\phi_{wire}$. Thus, the above constraint on the phase change can be rewritten as $\phi_{loop}(B)\!\equiv\! 2\phi_{cc}\!+\!2\phi_{wire}\!=\!2\pi n$. This equation is used to describe Little-Parks (LP) oscillations~\cite{Little:1962,Parks:1964} of various types~\cite{Belkin:2011,Hopkins:2005} that appear when $B$ is varied. The periodicity of these oscillations is related to the fact that whenever the magnetic field is such that $2\phi_{cc}(B)\!=\!2\pi k$ the family of free-energy minima is identical to $B\!=\!0$ case, if the vorticity is redefined as $n\!\to\!(n\!\pm\!k)$, where $k$ is an integer~\cite{Hopkins:2005}.
Sufficiently small thickness of the film allows us to neglect its effect on the external field.

Below, we present data from sample A, which incorporates two nanowires that are 200~nm long, 17~nm thick and 22-25~nm wide. The spacing between the nanowires is 13~$\mu$m. Data obtained from sample B, incorporating nanowires with other parameters, show similar behavior and are relegated to the Supplemental Material (SM-3,8). Measurements in the temperature range from 0.3~K to 3.0~K were performed in $^3$He system. The microwave signal from the output of Agilent N5320A vector network analyzer was directed to the input of the Fabry-P\'{e}rot resonator through a total of $\sim$47~dB of attenuators, thermally anchored at cryogenic temperatures. The sample was mounted inside a brass Faraday cage that screened external RF noise, but was transparent for DC magnetic field. The ground planes of the resonator were connected to the Faraday cage to achieve a superior thermalization of the sample. The signal from the output of the resonator traveled through a low temperature microwave amplifier, which had $\sim$40~dB gain and $\sim$2.6~K input noise temperature. To prevent this noise from impacting the sample, we placed two thermally anchored isolators between the sample and the amplifier. The total attenuation of the output line, excluding the amplifier, was $\sim$47~dB. Measurements in the temperature interval from 60~mK to 250~mK were carried out in a dilution refrigerator. Its microwave line was similar to that of $^3$He system, but had a higher attenuation ($\sim$54~dB). More detailed description of the experimental setup can be found in the Supplemental Material (SM-2).

\begin{figure}[htbp]
\centering
	\includegraphics[width=86mm]{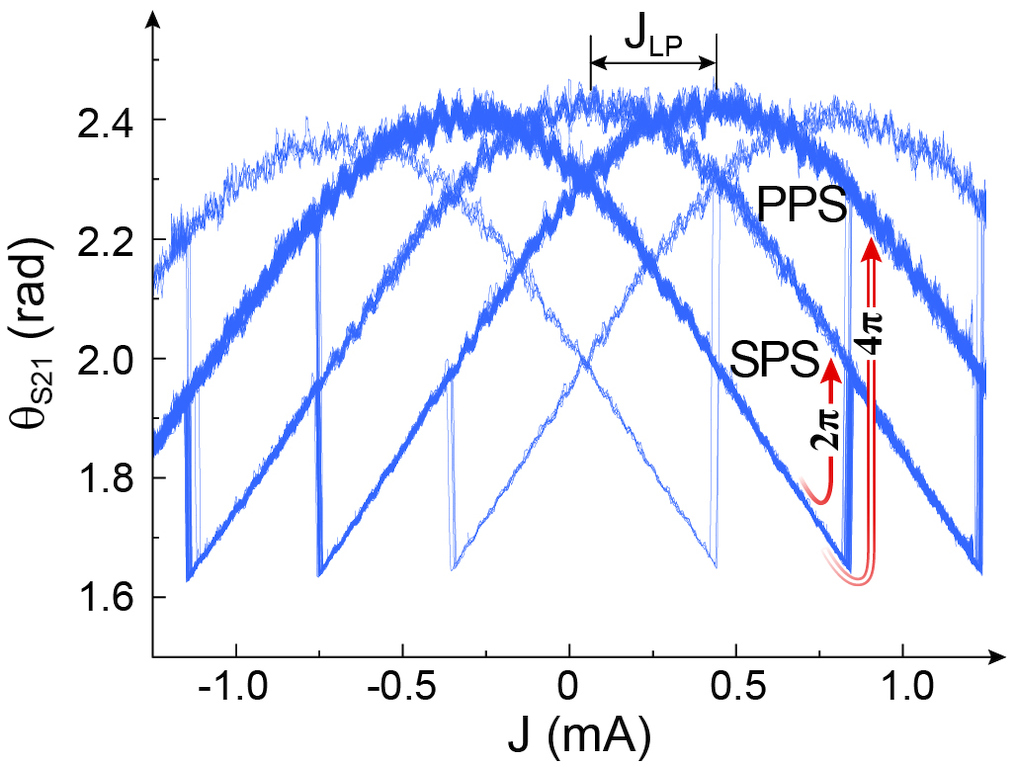}
	\caption{Dependence of the transmission phase on the current in the solenoid. The bath temperature is 350~mK. Periodically spaced ``parabolas'', each corresponding to a particular vorticity or the quantum winding number, $n$, are clearly observed. SPS result in jumps of $\theta_{S21}$ to the next parabola (single-line red arrow) and $n\!\to\!(n\!\pm\!1)$ change in the vorticity. PPS result in jumps of $\theta_{S21}$ to the next nearest parabolas (double-line red arrow) and $n\!\to\!(n\!\pm\!2)$ change in the vorticity. The Little-Parks period is marked as $J_{LP}$. Regions where switching events form clusters correspond to magnetic fields and vorticity values at which the current in the wires is near the critical current.} 
	\label{fig2}
\end{figure}

We detect phase slips by measuring the resonator's transmission phase shift, $\theta_{S21}$, defined as the difference between phases of the microwave signal at the output and the input of the resonator. The frequency of the input signal, $f_0$, is set equal to that of the resonator's fundamental mode at $B\!=\!0$ ($f_0\!\approx\! 5$~GHz for sample A). When the field changes, the resonance frequency shifts; consequently, the transmission through the resonator decreases. After cycling the magnetic field that is generated by current, $J$, in solenoid multiple times, we obtain the dependence shown in Fig.~\ref{fig2}. The transmission phase $\theta_{S21}(J)$ is a multivalued function composed of a periodic set of ``parabolas'' that are separated by the Little-Parks period, $J_{LP}$~\cite{Belkin:2011}. Each parabola corresponds to a state with particular vorticity. In what follows we will show that transitions between different parabolas below $\sim$0.9~K proceed through quantum tunneling. Every jump  to the nearest parabola is described by  $n\!\to\!(n\!\pm\!1)$ vorticity change and corresponds to an SPS event, which takes place in one of the wires~\cite{Little:1967}. Interestingly enough, apart from SPS we also detect $n\!\to\!(n\!\pm\!2)$ transitions for which  the phase jumps to the next nearest parabola. These experimentally detected jumps correspond to paired phase slips (PPS). Later we will argue that PPS can be due to both quantum and classical effects. Unlike previous experiments~\cite{Sahu:2009, Li:2011}, an important feature of the detection method employed here is that it does not rely on overheating of nanowires by phase slips.

\begin{figure}[htbp]
\centering
	\includegraphics[width=86mm]{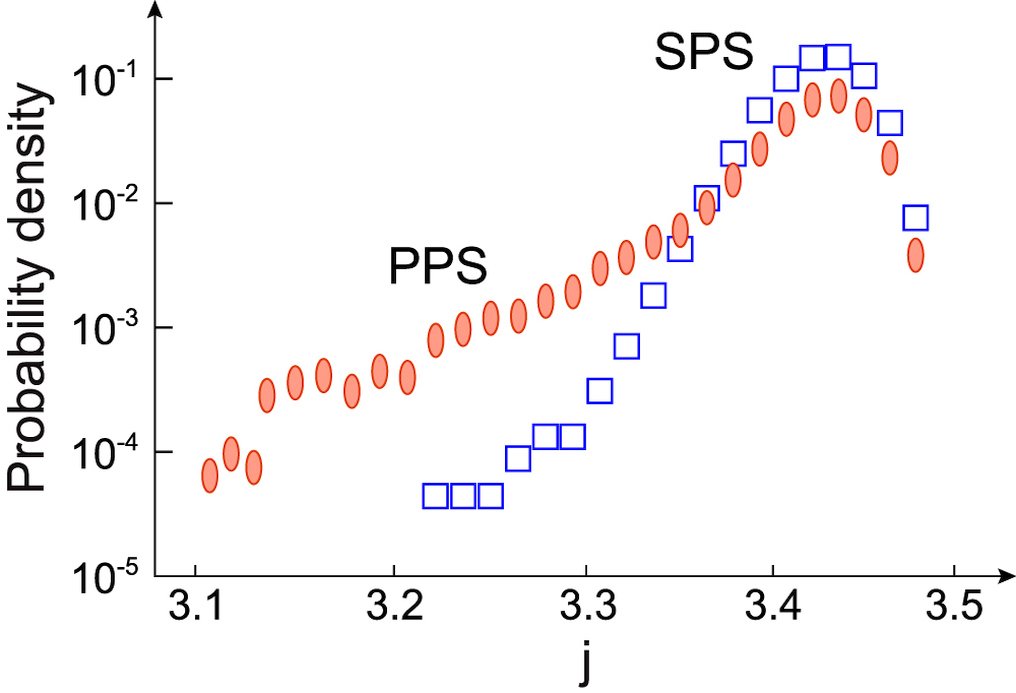}
	\caption{Probability densities of SPS and PPS as functions of normalized current in solenoid. Blue squares correspond to SPS; red ovals correspond to PPS transitions. These measurements have been performed at $T=60$~mK.} 
	\label{fig3}
\end{figure}

The switching of the loop vorticity has a stochastic nature, \ie\ in every measurement it occurs at a slightly different current $J$. To characterize the switching events, we employ a statistical approach. In particular, within each current interval where phase slips take place, we extract individual switching events, identify them as SPS or PPS and then compute their switching current distributions (see SM-5). One of the examples of the obtained probability densities as functions of the normalized current in the solenoid, $j\!=\!(J-J_0)/J_{LP}$, is shown in Fig.~\ref{fig3}. Here $J_0$ is the center of the corresponding parabola, i.e., its maximum. The chosen normalization reveals how far the system can be driven away from its equilibrium vorticity before a phase slip takes place. If the system were completely classical {\it and} followed its lowest energy state, then the first SPS would occur at $j\!=\!1/2$ (assuming $n\!=\!0$ initially) and PPS would never happen. According to the data, the phase slips take place at values higher than $j\!=\!3$ (see Fig.~\ref{fig3}), which is a metastable state quite far from the equilibrium. Therefore, it is clear that at $j\!=\!0$ and $n\!=\!0$ and even at $j\!=\!1/2$ and $n\!=\!0$ the barrier for phase slips is much higher than the scale of thermal or quantum fluctuations. Note that since on the timescale of our measurements the quantum tunneling of phase slips is not observed in the vicinity of parabolas' intersections ($j\!=\!\pm1/2, \pm3/2$ etc.) the so-called anticrossing, characteristic for qubits, does not show up in Fig.~\ref{fig2}.

\begin{figure}[htbp]
\centering
	\includegraphics[width=86mm]{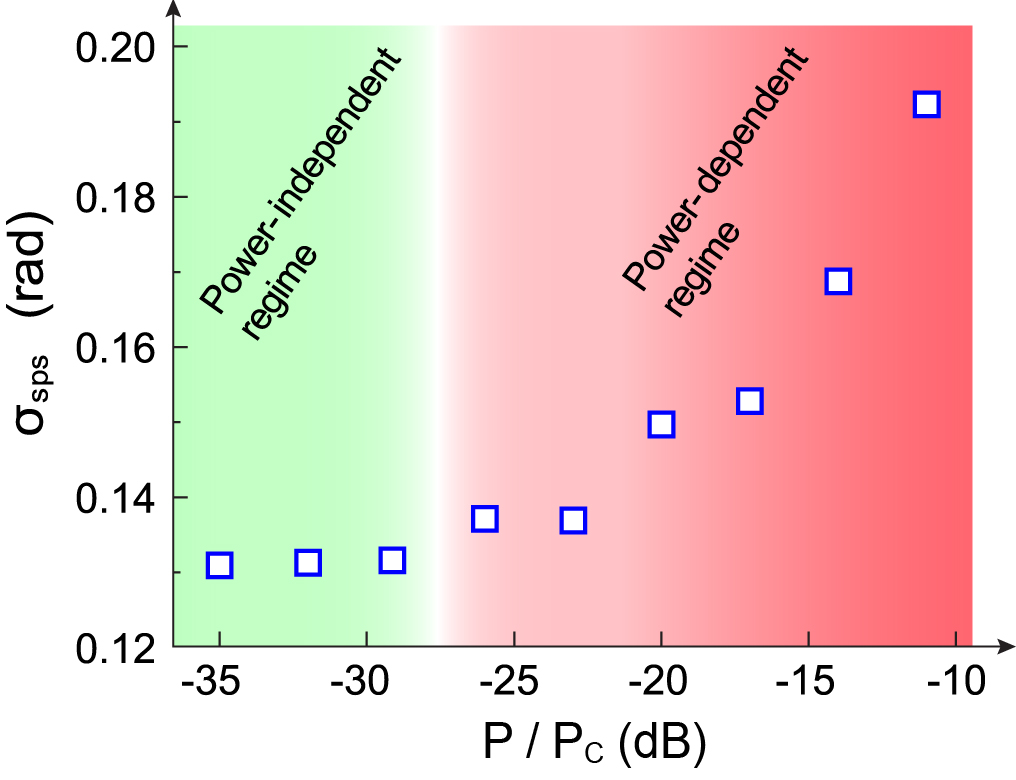}
	\caption{Dependence of the standard deviation of SPS switching current distribution on the reduced input power of the microwave signal. The measurements are made at 60~mK. $P_c$ corresponds to a power that induces in the wires a current equal to their critical current.}
	\label{fig4}
\end{figure}

Let us discuss now the choice of the measurement power. The current flowing in the wires includes two components: the screening current due to the magnetic field and the oscillating current due to the microwave signal. To study MQT of phase slips in nanowires it is imperative to choose such a power of the input signal, $P$, which would not stimulate additional phase slips, \ie\ would not increase their probability. The influence of the microwave radiation on the switching current distribution of SPS is shown in Fig.~\ref{fig4}. As one can see, at $P/P_c\!\gtrsim\!-\!27$~dB the width of SPS switching current distributions, \Ssps, starts to increase with the power $P$. Here the critical power, $P_c$, is a power of the radiation that induces oscillating currents with the amplitude equal to the critical superconducting current of the nanowires. This broadening of the distributions happens due to the fact that the microwave field can cause additional switching events. Yet, if $P/P_c\!\lesssim\!-\!27$~dB then \Ssps\ is independent of the power.  Therefore, to exclude the impact of radiation on the loop vorticity, all the measurements presented below are performed with the microwave power that is 30~dB less than $P_c$, \ie\ with the power that is about 1000 times weaker than the power needed to overcome the critical current in the wires. In the example of Fig.~\ref{fig4} the critical power is $P_c\!\approx\!-71$~dBm evaluated at the input of the resonator.

As one can see in Fig.~\ref{fig4}, we express the standard deviation in radians. This is done by dividing \Ssps, measured in Amps, by the Little-Parks period $J_{LP}$ and then multiplying the ratio by $2\pi$. Such normalization takes into account that the Little-Parks period corresponds to the $2\pi$ change of the phase difference accumulated on both wires.  

\begin{figure}[htbp]
\centering
	\includegraphics[width=86mm]{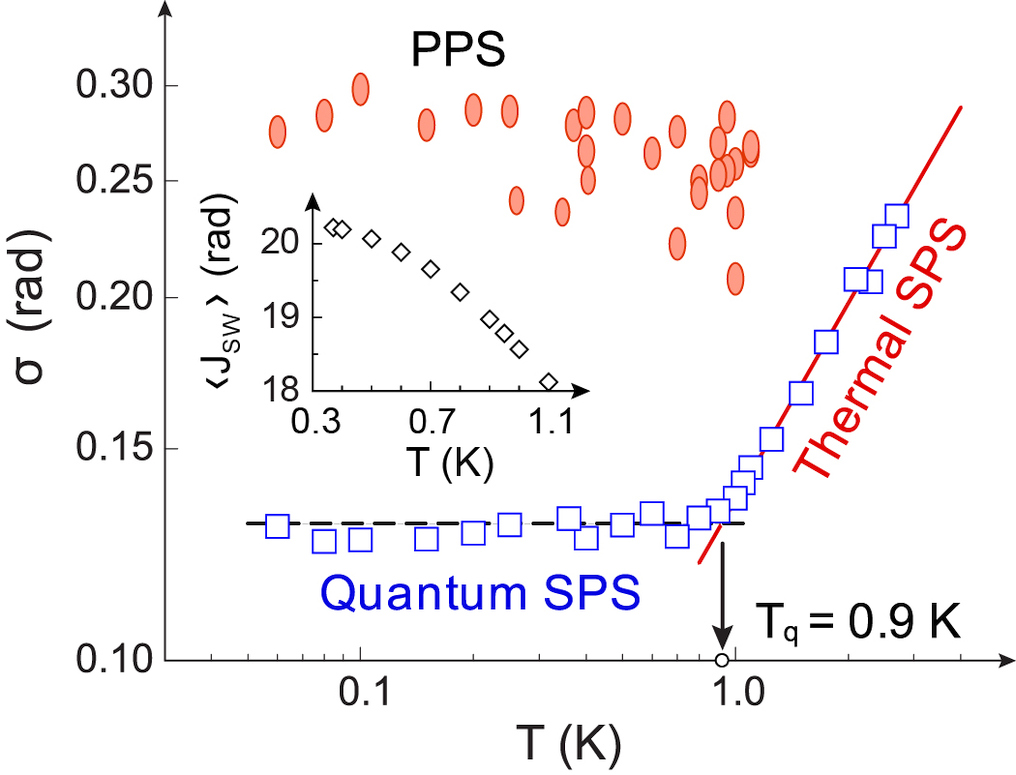}
	\caption{Temperature dependence of standard deviations of SPS and PPS switching current distributions. Temperature $T_q$ (black arrow) marks a crossover from thermal activation to quantum tunneling of SPS. The solid red line is the best fit of \Ssps$(T)$ in the thermally activated regime: $\sigma_{SPS}(T\!>\!T_q)\!=\!0.136\,T\,^{0.54}$\,rad. The dashed black line shows the best fit in the quantum tunneling regime: $\sigma_{PPS}(T\!<\!T_q)\!=\!0.130$\,rad. In the inset the average switching current versus temperature, $J_{sw}(T)$, is shown. The switching current continues to increase with cooling down to temperatures much lower than $T_{q}$. 
	}
	\label{fig5}
\end{figure}

From a series of switching current distributions measured at different temperatures, we compute standard deviations of the switching currents (see SM-5) and plot them in Fig.~\ref{fig5}. When the sample is cooled down \Ssps\ exhibits the expected behavior in accordance with the Kurkij\"{a}rvi power-law scaling~\cite{Kurkij:1972}, derived for thermally activated escape processes. The best fit is $\sigma_{SPS}(T)\!=\! 0.136\,T\,^{0.54}$, proving the thermally activated nature of the vorticity jumps. At low temperatures, we observe a reproducible saturation of \Ssps, which is explained in terms of macroscopic quantum tunneling from a state with higher vorticity to a state with lower vorticity. The temperature $T_q$ is understood as a crossover between the regimes of thermally activated phase slips ($T\!>\! T_q$) and quantum tunneling ($T\!<\!T_q$). 

There are two facts which provide strong evidence that the observed saturation of \Ssps\ is not a result of an imperfect cooling of the sample, but indeed is due to internal quantum fluctuations in the samples, \ie\ due to the macroscopic quantum tunneling. First, the obtained value of $T_q\!\approx\!0.90$~K is very close to 0.87~K, which is expected from the previously determined linear dependence~\cite{Aref:2012} of $T_q$ on the critical temperature $T_c$ (see SM-4). Note that this alternative estimate is based on a qualitatively different experimental technique, namely dc measurements of the switching current of single nanowires~\cite{Aref:2012}. Second, the average SPS switching current, $J_{sw}(T)$, continues to grow with cooling even when the standard deviation is already saturated (see inset in Fig.~\ref{fig5}). The latter fact appears to be in a good agreement with the analysis of the statics of the switching events, developed by Kurkijarvi~\cite{Kurkij:1972} and generalized by Garg~\cite{Garg:1995}. In particular, the observation of the increase of the switching current with cooling reflects the fact that the critical current of the nanowires is expected to grow with cooling even at $T\!\ll\!T_c$ according to Bardeen’s formula, which was discussed in, \eg, Ref.~\cite{Aref:2012}. Nonetheless, although we are convinced that the explanation of \Ssps\ saturation in terms of macroscopic quantum tunneling is the most natural one and we do not see any practical explanations in terms of spurious effects, it is impossible to prove with $100\%$ certainty that alternative explanations do not exist.

The dependence of the standard deviation of PPS switching current distribution, \Spps, on the temperature significantly differs from that of SPS. First, PPS are almost never observed in the thermal activation regime ($T\!>\!T_q$). 
Second, at temperatures where PPS do occur, their switching current distributions are broader than the distributions of SPS: $\sigma_{PPS}\!>\!\sigma_{SPS}$ (see also Fig.~\ref{fig3}). 

\begin{figure}[htbp]
\centering
	\includegraphics[width=86mm]{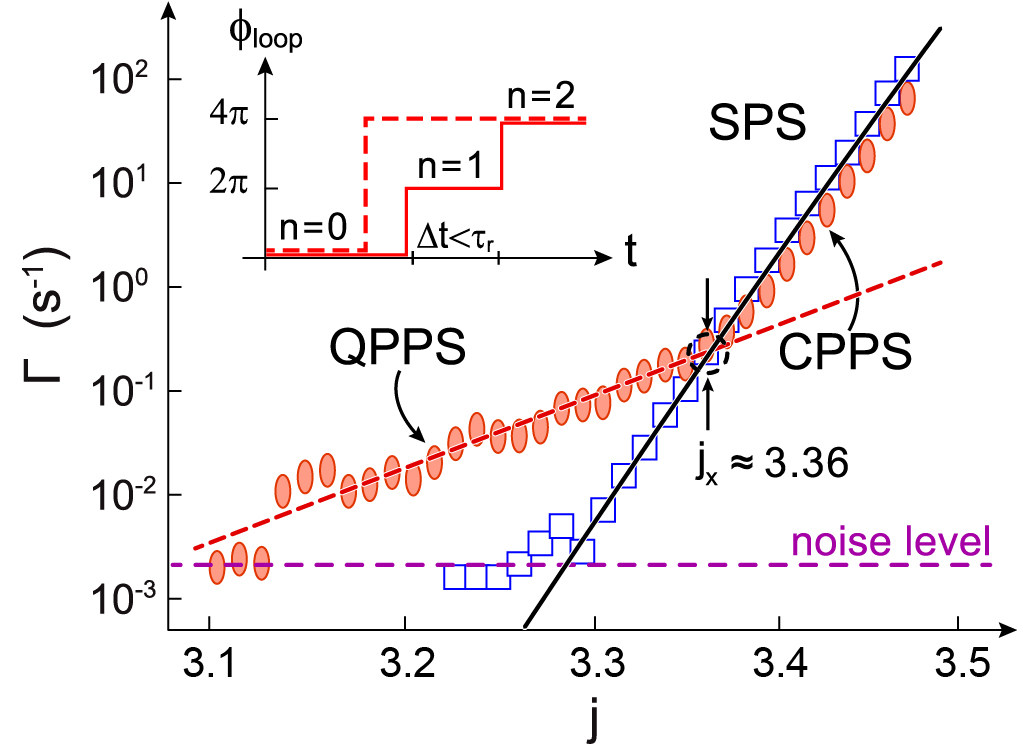}
	\caption{Rates of SPS (blue squares) and PPS (red ovals) as functions of the normalized current in the solenoid. The bath temperature is 60~mK. Single (blue squares) and paired (red ovals) phase slip rates are obtained from the switching current distributions shown in Fig.~\ref{fig3}. The solid black line is the Kurkij\"{a}rvi-Garg fit for the rate of SPS. The dashed red line is the QPPS fit, generated according to Korshunov's theory of QPPS~\cite{Korshunov:1987}. Rates of SPS and PPS are equal at the current $j_{x}$. The inset shows the schematic difference between CPPS (red solid line) and QPPS (red dashed line).}
	\label{fig6}
\end{figure}

To make the next step in our analysis, we introduce the phase slip rate, \G$(J)$, defined such that probability of a phase slip in the time interval $dt$ equals \G$dt$. Unlike the probability density, the rate \G$(J)$ does not depend on the sweep speed $v_J\!=\!dJ/dt$ (note that the maximum of the probability density shifts to higher currents if $v_{J}$ is increased.) In our experiments, the current in the solenoid was varied sinusoidally. The frequency of oscillations was 0.1 Hz, limited by the time resolution of the setup and the response time due to eddy currents induced in Faraday cage. According to Kurkij\"{a}rvi~\cite{Kurkij:1972}, the probability density of phase slips for a given current sweep speed has one--to--one correspondence to \G$(J)$. Following the method outlined in references~\cite{Kurkij:1972,Fulton:1974,Bezryadin:2012}, we performed a statistical analysis of more than $10^4$ switching events and computed the rates of SPS and PPS, denoted as \Gsps\ and \Gpps, respectively (see SM-5). The result of these calculations is presented in Fig.~\ref{fig6}.

An important conclusion following from Fig.~\ref{fig6} is that the rate of PPS exceeds the rate of SPS by approximately one order of magnitude at the lowest currents, at which the switching events have been observed. Demonstration of this regime constitutes the key finding of the present work. If the observed trend continues down to zero current, the rate of SPS is expected to be negligible at $J\!=\!0$ compared to the PPS rate.

One may argue that the observed $4\pi$ phase shifts are caused solely by two sequential $2\pi$ slips (schematically illustrated by the red solid line on the inset in Fig.~\ref{fig6}) crossing the loop. Let us assume that this is true. In that case every SPS decreases the supercurrent in the loop bringing the system from a highly metastable state closer to the equilibrium. The reduction of the supercurrent results in the increase of the energy barrier for the subsequent SPS~\cite{Garg:1995,Aref:2012}. At the same time, for both thermal activation ($T\!>\!T_q$) or quantum tunneling ($T\!<\!T_q$), each phase slip causes a release of energy $Q_1\!\approx\! I\Phi_0$~\cite{Anderson:1964,Sahu:2009,Shah:2008,Pekker:2009}, where $\Phi_0$ is the flux quantum and $I$ is the current in the nanowires. This energy heats up the wire and reduces the barrier. During the energy relaxation time, $\tau_r$, the wire stays ``warm'' and the second SPS can occur with a certain probability after a delay time, $\Delta t$. The larger the current is at which the first SPS takes place, the more energy it releases ($Q_1\!\sim\!I\!\sim\!j$), and therefore the higher the probability is that the second SPS will occur within the time $\tau_r$. Note that the delay $\Delta t$ is always smaller than $\tau_r$ because, as soon as the relaxation time has elapsed, the wire loop enters a new metastable state with a current significantly less than the critical one, and so it cannot switch.
The scale of $\tau_r$ is on the order of nanoseconds~\cite{Sahu:2009}. This is much shorter than the resolution of our setup, which is in the range of a few milliseconds. Therefore, we cannot directly distinguish each of these consecutive single phase slip jumps. As a consequence, all switching events are registered as single jumps, changing $\phi_{loop}$ either by $2\pi$ or $4\pi$. If the $4\pi$ jumps are composed out of two separate phase slips, the first being the cause for the second one, then we call them classically-paired phase slips (CPPS).

Recalling our assumption that there exist only SPS and sequential $4\pi$ phase slips, \ie\ CPPS, we conclude that the following equation must hold: \Psps+\Pcpps=1. Here \Psps\ and \Pcpps\ are the probabilities of SPS and CPPS, correspondingly. According to Fig.~\ref{fig6}, \Gsps$>$\Gcpps\ at $j\!>\!j_x$, where $j_x$ is the current at which \Gsps$(j_x)$=\Gcpps$(j_x)$. Therefore, the probabilities must satisfy the condition: \Psps$>$\Pcpps, bringing us to the conclusion that \Pcpps$(j\!>\!j_x)\!<$0.5. As we go from higher to lower switching currents, the amount of energy released as a result of each SPS event goes down because $Q_1\!\sim\!j$. This reduction should lower the probability that one SPS causes one more SPS within the energy relaxation time interval. Thus, the relative percentage of the CPPS should decline as the current in solenoid decreases. This conclusion obviously contradicts the experimental results (see Fig.~\ref{fig6}), which show that the rate of PPS is larger than SPS at relatively low values of the current in the solenoid. Consequently, the assumption that all PPS are simply two sequential SPS (\ie\ the assumption that each experimentally observed PPS represents CPPS) is incorrect and some other type of PPS must exist. (Let us note here, that the dominance of PPS over SPS at low currents became even slightly stronger when we improved the filtering of the signal lines and reduced the measurement power, indicating that the effect is not caused by an external noise, see SM-6 and Fig.~S3.)

We suggest that high rate of \Gpps\ at $j\!<\!j_x$ can be explained by the existence of a different type of phase slip pairs. We call them quantum-paired phase slips (QPPS). The theoretical background of QPPS was explained in Refs.~\cite{Guinea:1986,Korshunov:1987,Zaikin:1990}. The main difference between classical and quantum pairing of phase slips is that, unlike in CPPS, in QPPS the two phase slips occur simultaneously and do not cause one another. The reason for such a simultaneous process is to minimize the quantum action of the tunneling event (see SM-9). Since thermal fluctuations are negligible at $T\!<\! T_q$, it is justified to characterize QPPS as {\it cotunneling} of phase slips. 

As it follows from Fig.~\ref{fig6}, \Gpps\ is smaller than \Gsps, but they go almost parallel to each other at $j\!>\!j_x$. Since classical pairing assumes the stimulation of the second phase slip by the first one, the above observation that the rate versus current curves are parallel to each other allows us to conclude that CPPS dominate over QPPS in the high current region. On the contrary, in the low current region the SPS and PPS curves decouple from each other indicating that the cotunneling of phase slips (\ie\ QPPS) starts to prevail. In the logical analysis above we rely on the understanding that QPPS is a transition completely independent and different from SPS, while CPPS is a composed event, the first stage of which is an SPS. Therefore the rate of CPPS is proportional to the rate of SPS while the rate of QPPS is not a function of the rate of SPS. We focus our attention on QPPS because they preserve the parity of the winding number of the superconducting order parameter and therefore QPPS could lead to novel topologically protected qubits~\cite{Kitaev:2003,Gottesman:2001,Dou:2002,Dou:2012}.

An approximate description of short superconducting nanowires similar to ours~\cite{Aref:2012} can be given by the Stewart-McCumber model~\cite{Bezryadin:2012}. In the framework of this model, the phase difference represents the position of some effective ``phase particle'' in a tilted ``washboard'' potential. The effective mass $m\!=\!\hbar^2C/(2e)^2$ of the phase particle is proportional to the shunt capacitance, $C$, connected in parallel with a superconducting junction. The effect of inertia in this analogy is due to the fact that if the effective phase particle is moving, that means that the capacitor is charged and so it would keep its charge for some time, stimulating the phase to change further. As usual, $\hbar$ is the reduced Plank constant and $e$ is the electron charge. The friction experienced by the particle is described by the effective viscosity $\eta\!=\!\hbar^2 /4e^2 R_{n}$, inversely proportional to the shunt impedance, $R_{n}$. As is true in the mechanical analogies, the effect of viscosity is to compete with the effect of the inertia. In the considered case the presence of a normal shunt causes the shunting capacitor to discharge more rapidly, thus reducing the inertia associated with the charge accumulation on the capacitor. The character of the particle motion is determined by the ratio between $m$, $\eta$ and the potential barrier height~\cite{Fulton:1974} $U\!=\!(2\sqrt{2}/3)(\hbar I_c/e)(1\!-\!I_{sw}/I_c)^{3/2}$, where $I_{sw}$ is the current in the wire just before the switching event and $I_c$ is the wire critical current. Note that such barrier height occurs if a cosine-shaped potential energy landscape is tilted. Within this model the critical current represents such a tilt of the potential which is sufficient to eliminate all metastable minima. In underdamped regime ($mU\!\gg\!\eta^2$), the ``heavy'' particle moves in a slightly viscous environment. The viscosity in this case does not play a significant role and thus can be neglected. The main parameter influencing the phase slip rate at a particular bias current is the barrier height $U$. 

Let us demonstrate that SPS correspond to the {\it underdamped} motion of the effective phase particle discussed above. For this purpose, we need to know approximate values of the effective shunt capacitance, the shunt impedance, and the critical current of the nanowires. According to previous studies~\cite{Brenner:2012,Aref:2012}, the critical current of our nanowires is of the order of $10\,\mu$A. The effective shunt impedance is set by the waveguide impedance of the resonator, which is $R_{n}\!\sim\! 50\, \Omega $. The estimate of the effective capacitance (the c.c.~strips) can be made based on the resonance frequency, $f_0$, of the resonator and the kinetic inductance of the wires, $L_k$: $C_1\!=\! \left(4\pi^2 f_0^2 L_k\right)^{-1}$. The kinetic inductance scales linearly with the wire length, $l$, and is inversely proportional to the the critical current~\cite{Bezryadin:2012,Likharev:1979}: $L_k\!\approx\! \left(2/{3\sqrt{3}}\right)(l/\xi)(\Phi_0/2\pi)(1/I_c)$, where $\xi$ is superconducting coherence length. Substituting $l\!=\!200$\,nm, $\xi\!\approx\!7$\,nm~\cite{Rogachev:2005}, $I_c\!=\!10\,\mu$A and $f_0\!\approx\!5$\,GHz, we find $L_k\!\approx\!3.5\!\times\!10^{-10}$\,H and $C_1\!\approx\! 3\!\times\! 10^{-12}\,$F. Thus, we arrive at the following estimates: $m_1\!\approx\!3\!\times\! 10^{-43}$ J$\cdot$s$^2$, $\eta_1\! \approx\!2\!\times\! 10^{-33}$\,J$\cdot $s, $U\!\approx\!2\!\times\!10^{-22}$\,J (here we use $I_{sw}/I_c\!\approx\!0.9$, which correlates well with the subsequent fitting analysis and is consistent with previous experiments~\cite{Aref:2012}). Such parameters indeed correspond to the underdamped regime, $m_1U\!\gg\!\eta_1^2$. Consequently, SPS can be analyzed in the framework of the underdamped Kurkij\"{a}rvi-Garg (KG) theory~\cite{Kurkij:1972,Garg:1995}. Following Aref {\it et al.}~\cite{Aref:2012}, we fit the rate of SPS in amorphous MoGe wires using the equation:
\begin{equation}
\label{eq:Gamma1}
{\it\Gamma}_{SPS}(j) = \Omega \, \exp\left[-\left(2\sqrt{2} \hbar I_c /3e k_B T_q\right)\left(1-j/j_c\right)^{3/2}\right]
\end{equation}

Here $\Omega$ is the phase slip attempt frequency, which is assumed constant, $k_{B}$ is the Boltzmann constant, $j_c$ is the normalized critical current in the solenoid, defined as $j_c\!=\!J_c/J_{LP}\!=\!I_c/I_{LP}$. Note that $J_c$ is the current in the solenoid that induces the Meissner current, entering the nanowires, that is equal to their critical current $I_c$. Current $I_{LP}$ is the Little-Parks period measured in the units of the current in the nanowires. The best fit generated by the KG formula with the use of three fitting parameters is shown in Fig.~\ref{fig6} (black solid line). It corresponds to $\Omega\!=\!0.9 \!\sim\! 10^{11}$\,Hz, $I_c\!=\!7.8\,\mu$A, $j_c\!=\!4.04$. The best fit values of $\Omega$ and $I_c$ are somewhat different from the previously reported values~\cite{Aref:2012} but they are of the same order of magnitude. The $j_c$ is close to, but higher than the experimentally observed maximum switching current $j_{max}\!\sim\!3.5$ (see Fig.~\ref{fig6}), as it should be. Thus, the results for SPS are in good agreement with the indirect SPS observations, performed with dc measurements~\cite{Aref:2012}. 

It is beneficial to note that the obtained value of $j_c$ may be utilized to find the coherence length, $\xi$, in superconducting wires. Indeed, from the Gor'kov-Josephson phase-evolution equation $d\phi_{wire}/dt\!=\!2eV/\hbar$, where voltage, $V$, between the ends of the wire depends on the kinetic inductance, $L_k$, as $V\!=\!L_k(dI/dt)$, one can obtain the phase difference between the opposite ends of the wires as function of the supercurrent~\cite{Likharev:1979}: $\phi_{wire}\!=\!\left(2/{3\sqrt{3}}\right)(l/\xi)(I/I_c)$ (the expression for $L_k$ was given above). The phase slips take place at $I\!\approx\! I_c$ when $\phi_c\! \equiv\! \phi_{wire}(I\!=\!I_c)\!\approx\! 2\pi j_c/2$. It is implied here that the phase accumulated on two wires right before the transition, $2\pi j_c$, is divided equally between them. The above formula for $\phi_c$ follows from the model of Hopkins {\it et al.}~\cite{Hopkins:2005}, according to which $\phi_{wire}(B,n)\!=\!\pi n\!-\!\phi_{cc}(B)$ and the phase drop on both electrodes depends on the magnetic field as $\phi_{cc}(B)\!=\!\pi(8G/\pi^2)(w_{cc}w_{wire}B/\Phi_0)$, and it is independent of the loop vorticity $n$. Here $w_{cc}$ is the width of the c.c.~strip ($\sim\!20\,\mu$m), $w_{wire}$ -- the spacing between the wires ($\sim\!13\,\mu$m for sample~A), $G\!\approx\!0.916$ -- the Catalan number, defined as $G\!=\!\sum_{n=0}^{\infty} (-1)^{n}/(2n+1)^2$. Thus, the estimate of the coherence length in the wires is $\xi\!=\!2l/\left(3\sqrt{3}\pi j_c\right)\!\approx\! 6.1$\,nm, which agrees well with independently reported values~\cite{White:1993,Rogachev:2005}.

We now turn to the discussion and analysis of PPS structure and characteristics. A pair of fluxoids moving in opposite directions and crossing {\it both wires simultaneously} (see Fig.~\ref{fig1}) does not charge the capacitor formed by the c.c.~strips. The effective mass of the phase particle in this process is much lower compared to the case when one or two phase slips occur on just one wire, charging the capacitor formed by the two halves of the c.c.~strip. Thus, such mechanism (Fig.~\ref{fig1}) of phase slip cotunneling is effectively less ``heavy'' and therefore is {\it much more probable} than the one, in which a fluxoid pair crosses a single wire. Due to the specifics of the fabrication process, our wires are not completely identical. The role of asymmetry is not well understood (future experiments will address the role of asymmetry by making samples asymmetric on purpose). Nonetheless, since QPPS have been observed in two samples, we conclude that the small degree of asymmetry, which can be accidentally present in real samples, does not play a crucial role in our experiment. Because the effective capacitance is much smaller for QPPS than for SPS, the effective mass of QPPS particle is also much smaller. As a result, unlike in the case of SPS, the influence of dissipation~\cite{Caldeira:1981} on the rate of QPPS is important and cannot be neglected. Indeed, as we will now show, a good fit to the data is obtained upon assumption that QPPS events are in the overdamped regime. Starting from this assumption, the QPPS rate, \Gqpps, can be roughly estimated by using Korshunov's instanton solution~\cite{Korshunov:1987}, which is schematically illustrated in Fig.~\ref{fig6}~(inset) by the dashed line. The original solution was obtained for the junctions with sinusoidal current-phase relationship (CPR). Since our nanowires possess a different CPR, we generalize the Korshunov instanton by including an additional parameter $\alpha$ characterizing the peculiarities of the spatio-temporal shape of the phase slips in superconducting wires. Our final expression for the QPPS rate is 
\begin{equation}
\label{eq:Gamma2}
{\it\Gamma}_{QPPS}=B_2 (j/j_c)^{-2/3} \exp[A_2 (j/j_c)^{2/3}] 
\end{equation}

Here $A_2$ and $B_2$ are defined through the critical current $I_c$, the effective viscosity $\eta_2$, the effective capacitance $C_2$, and the quantum resistance $R_q\!=\!\pi\hbar /2e^2$ as: $A_2\!=\!12 \pi(\alpha \eta_2 R_q C_2 I_c/4\pi\hbar e)^{1/3}$,

\noindent 
$B_2\!=\!(12\pi/A_2)(\alpha \eta_2/\hbar)^{1/2}(I_c/2e)\exp[-8\pi\{\eta_2/\hbar+[2/(\alpha \eta_2/\hbar)^{1/2}](A_2/12\pi)^{3/2}\}]$. From the best fit for the rate of QPPS (red dashed line in Fig.~\ref{fig6}) we determine $A_2\!=\!91.4$, $B_2\!=\!1.5 \cdot 10^{-36}\,$s$^{-1}$. The value of the normalized critical current is not independent here, but is set by the SPS fit, \ie\ $j_c\!=\!4.04$.	The theory is valid in the limit of large viscosity, \ie, in the overdamped regime, when $\beta\!\equiv\!\eta_2/(\hbar^2 R_{q}C_2 I_{c} /4\pi e)^{1/2}\!\gg\! 1$. This condition is satisfied by choosing $\alpha\!=\!6$, $C_2\!=\!3.1 \times 10^{-17}$\,F, $\eta_2\!=\!2.8\times 10^{-34}$\,J$\cdot$\,s, which follow from the best fit values of $A_2$ and $B_2$ and deviate minimally from their expected values (see SM-7). Here $\eta_2$ is the effective viscosity for the QPPS events. The fitting parameters are in agreement with the assumption that QPPS corresponds to the tunneling in the overdamped regime and they correlate well with the fitting parameters for the sample B (see SM-8).

The fits generated by the Korshunov model are in excellent agreement with the data. However, it should be noted that the model is constructed for the case of sinusoidal dissipation, characteristic to normal shunts with pronounced charge discreteness, such as normal metal tunnel junctions. Because of that, the dissipative part of the quantum action contains a squared sinusoidal term (see equation (3.61) in Ref.~\cite{Zaikin:1990} and SM-9). The charge-discrete transport is not an unexpected phenomenon in nanowires. It has been theoretically predicted~\cite{Nazarov:1999, Golubev:2001} and experimentally observed on MoGe nanowires~\cite{Bollinger:2006}.
	
Qualitatively speaking, the higher rate of QPPS compared to SPS is explained by three facts: (1) The QPPS are effectively ``lighter'' since the net voltage they generate on each c.c.~strip is zero and thus the capacitive ``inertia'' is weak for them. (2) In the case QPPS the contribution of the environment to the quantum action is much smaller compared to single quantum phase slips because of the $4\pi$ periodicity of the action, as per the Guinea-Sch\"{o}n-Korshunov argument (see SM-9). (3) The electromagnetic emission generated by QPPS into the resonator, which is a dissipative effect slowing down macroscopic quantum tunneling~\cite{Caldeira:1981}, is reduced, again because the pair of phase slips does not produce any net ac voltage on the center conductor electrode of the coplanar waveguide resonator. 

In conclusion, we study the stability of fluxoid states in superconducting loops using microwave measurements. We provide direct evidence that at low temperatures the change of the loop vorticity is realized by macroscopic quantum tunneling of individual phase slips through nanowires forming the loop. We discover that if the bias is sufficiently low then cotunneling of two phase slips, \ie\ a quantum-paired phase slip, is exponentially more likely to occur than a single phase slip. Our future goal will be to use such parity conserving macroscopic tunneling for building parity protected qubits. 
	
We thank Yu. V. Nazarov, N. Prokofiev and A. D. Zaikin for helpful discussions. The work was supported by the DOE Award No. DE-FG0207ER46453, and by the NSF grants No. DMR10-05645 and No. ECCS-1408558.

\bibliography{Bibliography}

\begin{thebibliography}{44}%
\makeatletter
\providecommand \@ifxundefined [1]{%
 \@ifx{#1\undefined}
}%
\providecommand \@ifnum [1]{%
 \ifnum #1\expandafter \@firstoftwo
 \else \expandafter \@secondoftwo
 \fi
}%
\providecommand \@ifx [1]{%
 \ifx #1\expandafter \@firstoftwo
 \else \expandafter \@secondoftwo
 \fi
}%
\providecommand \natexlab [1]{#1}%
\providecommand \enquote  [1]{``#1''}%
\providecommand \bibnamefont  [1]{#1}%
\providecommand \bibfnamefont [1]{#1}%
\providecommand \citenamefont [1]{#1}%
\providecommand \href@noop [0]{\@secondoftwo}%
\providecommand \href [0]{\begingroup \@sanitize@url \@href}%
\providecommand \@href[1]{\@@startlink{#1}\@@href}%
\providecommand \@@href[1]{\endgroup#1\@@endlink}%
\providecommand \@sanitize@url [0]{\catcode `\\12\catcode `\$12\catcode
  `\&12\catcode `\#12\catcode `\^12\catcode `\_12\catcode `\%12\relax}%
\providecommand \@@startlink[1]{}%
\providecommand \@@endlink[0]{}%
\providecommand \url  [0]{\begingroup\@sanitize@url \@url }%
\providecommand \@url [1]{\endgroup\@href {#1}{\urlprefix }}%
\providecommand \urlprefix  [0]{URL }%
\providecommand \Eprint [0]{\href }%
\providecommand \doibase [0]{http://dx.doi.org/}%
\providecommand \selectlanguage [0]{\@gobble}%
\providecommand \bibinfo  [0]{\@secondoftwo}%
\providecommand \bibfield  [0]{\@secondoftwo}%
\providecommand \translation [1]{[#1]}%
\providecommand \BibitemOpen [0]{}%
\providecommand \bibitemStop [0]{}%
\providecommand \bibitemNoStop [0]{.\EOS\space}%
\providecommand \EOS [0]{\spacefactor3000\relax}%
\providecommand \BibitemShut  [1]{\csname bibitem#1\endcsname}%
\let\auto@bib@innerbib\@empty
\bibitem [{\citenamefont {Leggett}(1980)}]{Leggett:1980}%
  \BibitemOpen
  \bibfield  {author} {\bibinfo {author} {\bibfnamefont {A.~J.}\ \bibnamefont
  {Leggett}},\ }\bibfield  {title} {\enquote {\bibinfo {title} {Macroscopic
  quantum systems and the quantum theory of measurement},}\ }\href {\doibase
  10.1143/PTPS.69.80} {\bibfield  {journal} {\bibinfo  {journal} {Prog.\
  Theor.\ Phys.\ Suppl.}\ }\textbf {\bibinfo {volume} {69}},\ \bibinfo {pages}
  {80--100} (\bibinfo {year} {1980})}\BibitemShut {NoStop}%
\bibitem [{\citenamefont {Lee}\ and\ \citenamefont {Jeong}(2011)}]{Lee:2011}%
  \BibitemOpen
  \bibfield  {author} {\bibinfo {author} {\bibfnamefont {C.~W.}\ \bibnamefont
  {Lee}}\ and\ \bibinfo {author} {\bibfnamefont {H.}~\bibnamefont {Jeong}},\
  }\bibfield  {title} {\enquote {\bibinfo {title} {Quantification of
  macroscopic quantum superpositions within phase space},}\ }\href {\doibase
  10.1103/PhysRevLett.106.220401} {\bibfield  {journal} {\bibinfo  {journal}
  {Phys.\ Rev.\ Lett.}\ }\textbf {\bibinfo {volume} {106}},\ \bibinfo {pages}
  {220401} (\bibinfo {year} {2011})}\BibitemShut {NoStop}%
\bibitem [{\citenamefont {Clarke}\ \emph {et~al.}(1988)\citenamefont {Clarke},
  \citenamefont {Cleland}, \citenamefont {Devoret}, \citenamefont {Esteve},\
  and\ \citenamefont {Martinis}}]{Clarke:1988}%
  \BibitemOpen
  \bibfield  {author} {\bibinfo {author} {\bibfnamefont {J.}~\bibnamefont
  {Clarke}}, \bibinfo {author} {\bibfnamefont {A.~N.}\ \bibnamefont {Cleland}},
  \bibinfo {author} {\bibfnamefont {M.~H.}\ \bibnamefont {Devoret}}, \bibinfo
  {author} {\bibfnamefont {D.}~\bibnamefont {Esteve}}, \ and\ \bibinfo {author}
  {\bibfnamefont {J.~M.}\ \bibnamefont {Martinis}},\ }\bibfield  {title}
  {\enquote {\bibinfo {title} {Quantum mechanics of a macroscopic variable: The
  phase difference of a josephson junction},}\ }\href {\doibase
  10.1126/science.239.4843.992} {\bibfield  {journal} {\bibinfo  {journal}
  {Science}\ }\textbf {\bibinfo {volume} {239}},\ \bibinfo {pages} {992--997}
  (\bibinfo {year} {1988})}\BibitemShut {NoStop}%
\bibitem [{\citenamefont {Jackel}\ \emph {et~al.}(1981)\citenamefont {Jackel},
  \citenamefont {Gordon}, \citenamefont {Hu}, \citenamefont {Howard},
  \citenamefont {Fetter}, \citenamefont {Tennant}, \citenamefont {Epworth},\
  and\ \citenamefont {Kurkij\"arvi}}]{Jackel:1981}%
  \BibitemOpen
  \bibfield  {author} {\bibinfo {author} {\bibfnamefont {L.~D.}\ \bibnamefont
  {Jackel}}, \bibinfo {author} {\bibfnamefont {J.~P.}\ \bibnamefont {Gordon}},
  \bibinfo {author} {\bibfnamefont {E.~L.}\ \bibnamefont {Hu}}, \bibinfo
  {author} {\bibfnamefont {R.~E.}\ \bibnamefont {Howard}}, \bibinfo {author}
  {\bibfnamefont {L.~A.}\ \bibnamefont {Fetter}}, \bibinfo {author}
  {\bibfnamefont {D.~M.}\ \bibnamefont {Tennant}}, \bibinfo {author}
  {\bibfnamefont {R.~W.}\ \bibnamefont {Epworth}}, \ and\ \bibinfo {author}
  {\bibfnamefont {J.}~\bibnamefont {Kurkij\"arvi}},\ }\bibfield  {title}
  {\enquote {\bibinfo {title} {Decay of the zero-voltage state in small-area,
  high-current-density josephson junctions},}\ }\href {\doibase
  10.1103/PhysRevLett.47.697} {\bibfield  {journal} {\bibinfo  {journal}
  {Phys.\ Rev.\ Lett.}\ }\textbf {\bibinfo {volume} {47}},\ \bibinfo {pages}
  {697--700} (\bibinfo {year} {1981})}\BibitemShut {NoStop}%
\bibitem [{\citenamefont {Wernsdorfer}\ \emph {et~al.}(1997)\citenamefont
  {Wernsdorfer}, \citenamefont {Bonet~Orozco}, \citenamefont {Hasselbach},
  \citenamefont {Benoit}, \citenamefont {Mailly}, \citenamefont {Kubo},
  \citenamefont {Nakano},\ and\ \citenamefont {Barbara}}]{Wernsdorfer:1997}%
  \BibitemOpen
  \bibfield  {author} {\bibinfo {author} {\bibfnamefont {W.}~\bibnamefont
  {Wernsdorfer}}, \bibinfo {author} {\bibfnamefont {E.}~\bibnamefont
  {Bonet~Orozco}}, \bibinfo {author} {\bibfnamefont {K.}~\bibnamefont
  {Hasselbach}}, \bibinfo {author} {\bibfnamefont {A.}~\bibnamefont {Benoit}},
  \bibinfo {author} {\bibfnamefont {D.}~\bibnamefont {Mailly}}, \bibinfo
  {author} {\bibfnamefont {O.}~\bibnamefont {Kubo}}, \bibinfo {author}
  {\bibfnamefont {H.}~\bibnamefont {Nakano}}, \ and\ \bibinfo {author}
  {\bibfnamefont {B.}~\bibnamefont {Barbara}},\ }\bibfield  {title} {\enquote
  {\bibinfo {title} {Macroscopic quantum tunneling of magnetization of single
  ferrimagnetic nanoparticles of barium ferrite},}\ }\href {\doibase
  10.1103/PhysRevLett.79.4014} {\bibfield  {journal} {\bibinfo  {journal}
  {Phys.\ Rev.\ Lett.}\ }\textbf {\bibinfo {volume} {79}},\ \bibinfo {pages}
  {4014--4017} (\bibinfo {year} {1997})}\BibitemShut {NoStop}%
\bibitem [{\citenamefont {Shor}(1995)}]{Shor:1995}%
  \BibitemOpen
  \bibfield  {author} {\bibinfo {author} {\bibfnamefont {P.~W.}\ \bibnamefont
  {Shor}},\ }\bibfield  {title} {\enquote {\bibinfo {title} {Scheme for
  reducing decoherence in quantum computer memory},}\ }\href {\doibase
  10.1103/PhysRevA.52.R2493} {\bibfield  {journal} {\bibinfo  {journal} {Phys.\
  Rev.~A}\ }\textbf {\bibinfo {volume} {52}},\ \bibinfo {pages} {R2493--R2496}
  (\bibinfo {year} {1995})}\BibitemShut {NoStop}%
\bibitem [{\citenamefont {Kitaev}(2003)}]{Kitaev:2003}%
  \BibitemOpen
  \bibfield  {author} {\bibinfo {author} {\bibfnamefont {A.~Yu.}\ \bibnamefont
  {Kitaev}},\ }\bibfield  {title} {\enquote {\bibinfo {title} {Fault-tolerant
  quantum computation by anyons},}\ }\href {\doibase
  10.1016/S0003-4916(02)00018-0} {\bibfield  {journal} {\bibinfo  {journal}
  {Ann.\ Phys.}\ }\textbf {\bibinfo {volume} {303}},\ \bibinfo {pages} {2--30}
  (\bibinfo {year} {2003})}\BibitemShut {NoStop}%
\bibitem [{\citenamefont {Gottesman}\ \emph {et~al.}(2001)\citenamefont
  {Gottesman}, \citenamefont {Kitaev},\ and\ \citenamefont
  {Preskill}}]{Gottesman:2001}%
  \BibitemOpen
  \bibfield  {author} {\bibinfo {author} {\bibfnamefont {D.}~\bibnamefont
  {Gottesman}}, \bibinfo {author} {\bibfnamefont {A.~Yu.}\ \bibnamefont
  {Kitaev}}, \ and\ \bibinfo {author} {\bibfnamefont {J.}~\bibnamefont
  {Preskill}},\ }\bibfield  {title} {\enquote {\bibinfo {title} {Encoding a
  qubit in an oscillator},}\ }\href {\doibase 10.1103/PhysRevA.64.012310}
  {\bibfield  {journal} {\bibinfo  {journal} {Phys.\ Rev.~A}\ }\textbf
  {\bibinfo {volume} {64}},\ \bibinfo {pages} {012310} (\bibinfo {year}
  {2001})}\BibitemShut {NoStop}%
\bibitem [{\citenamefont {Dou\ifmmode~\mbox{\c{c}}\else \c{c}\fi{}ot}\ and\
  \citenamefont {Vidal}(2002)}]{Dou:2002}%
  \BibitemOpen
  \bibfield  {author} {\bibinfo {author} {\bibfnamefont {B.}~\bibnamefont
  {Dou\ifmmode~\mbox{\c{c}}\else \c{c}\fi{}ot}}\ and\ \bibinfo {author}
  {\bibfnamefont {J.}~\bibnamefont {Vidal}},\ }\bibfield  {title} {\enquote
  {\bibinfo {title} {Pairing of cooper pairs in a fully frustrated
  josephson-junction chain},}\ }\href {\doibase 10.1103/PhysRevLett.88.227005}
  {\bibfield  {journal} {\bibinfo  {journal} {Phys.\ Rev.\ Lett.}\ }\textbf
  {\bibinfo {volume} {88}},\ \bibinfo {pages} {227005} (\bibinfo {year}
  {2002})}\BibitemShut {NoStop}%
\bibitem [{\citenamefont {Gladchenko}\ \emph {et~al.}(2009)\citenamefont
  {Gladchenko}, \citenamefont {Olaya}, \citenamefont {Dupont-Ferrier},
  \citenamefont {Dou\ifmmode~\mbox{\c{c}}\else \c{c}\fi{}ot}, \citenamefont
  {Ioffe},\ and\ \citenamefont {Gershenson}}]{Gladchenko:2009}%
  \BibitemOpen
  \bibfield  {author} {\bibinfo {author} {\bibfnamefont {S.}~\bibnamefont
  {Gladchenko}}, \bibinfo {author} {\bibfnamefont {D.}~\bibnamefont {Olaya}},
  \bibinfo {author} {\bibfnamefont {E.}~\bibnamefont {Dupont-Ferrier}},
  \bibinfo {author} {\bibfnamefont {B.}~\bibnamefont
  {Dou\ifmmode~\mbox{\c{c}}\else \c{c}\fi{}ot}}, \bibinfo {author}
  {\bibfnamefont {L.~B.}\ \bibnamefont {Ioffe}}, \ and\ \bibinfo {author}
  {\bibfnamefont {M.~E.}\ \bibnamefont {Gershenson}},\ }\bibfield  {title}
  {\enquote {\bibinfo {title} {Superconducting nanocircuits for topologically
  protected qubits},}\ }\href@noop {} {\bibfield  {journal} {\bibinfo
  {journal} {Nat.\ Phys.}\ }\textbf {\bibinfo {volume} {5}},\ \bibinfo {pages}
  {48--53} (\bibinfo {year} {2009})}\BibitemShut {NoStop}%
\bibitem [{\citenamefont {Bell}\ \emph {et~al.}(2014)\citenamefont {Bell},
  \citenamefont {Paramanandam}, \citenamefont {Ioffe},\ and\ \citenamefont
  {Gershenson}}]{Bell:2014}%
  \BibitemOpen
  \bibfield  {author} {\bibinfo {author} {\bibfnamefont {M.~T.}\ \bibnamefont
  {Bell}}, \bibinfo {author} {\bibfnamefont {J.}~\bibnamefont {Paramanandam}},
  \bibinfo {author} {\bibfnamefont {L.~B.}\ \bibnamefont {Ioffe}}, \ and\
  \bibinfo {author} {\bibfnamefont {M.~E.}\ \bibnamefont {Gershenson}},\
  }\bibfield  {title} {\enquote {\bibinfo {title} {Protected josephson rhombus
  chains},}\ }\href {\doibase 10.1103/PhysRevLett.112.167001} {\bibfield
  {journal} {\bibinfo  {journal} {Phys.\ Rev.\ Lett.}\ }\textbf {\bibinfo
  {volume} {112}},\ \bibinfo {pages} {167001} (\bibinfo {year}
  {2014})}\BibitemShut {NoStop}%
\bibitem [{\citenamefont {Dou\ifmmode~\mbox{\c{c}}\else \c{c}\fi{}ot}\ and\
  \citenamefont {Ioffe}(2012)}]{Dou:2012}%
  \BibitemOpen
  \bibfield  {author} {\bibinfo {author} {\bibfnamefont {B.}~\bibnamefont
  {Dou\ifmmode~\mbox{\c{c}}\else \c{c}\fi{}ot}}\ and\ \bibinfo {author}
  {\bibfnamefont {L.~B.}\ \bibnamefont {Ioffe}},\ }\bibfield  {title} {\enquote
  {\bibinfo {title} {Physical implementation of protected qubits},}\
  }\href@noop {} {\bibfield  {journal} {\bibinfo  {journal} {Rep.\ Prog.\
  Phys.}\ }\textbf {\bibinfo {volume} {75}},\ \bibinfo {pages} {072001}
  (\bibinfo {year} {2012})}\BibitemShut {NoStop}%
\bibitem [{\citenamefont {Guinea}\ and\ \citenamefont
  {Sch\"{o}n}(1986)}]{Guinea:1986}%
  \BibitemOpen
  \bibfield  {author} {\bibinfo {author} {\bibfnamefont {F.}~\bibnamefont
  {Guinea}}\ and\ \bibinfo {author} {\bibfnamefont {G.}~\bibnamefont
  {Sch\"{o}n}},\ }\bibfield  {title} {\enquote {\bibinfo {title} {Coherent
  charge oscillations in tunnel junctions},}\ }\href@noop {} {\bibfield
  {journal} {\bibinfo  {journal} {Europhys.\ Lett.}\ }\textbf {\bibinfo
  {volume} {1}},\ \bibinfo {pages} {585} (\bibinfo {year} {1986})}\BibitemShut
  {NoStop}%
\bibitem [{\citenamefont {Korshunov}(1987)}]{Korshunov:1987}%
  \BibitemOpen
  \bibfield  {author} {\bibinfo {author} {\bibfnamefont {S.~E.}\ \bibnamefont
  {Korshunov}},\ }\bibfield  {title} {\enquote {\bibinfo {title} {Coherent and
  incoherent tunneling in a josephson junction with a``periodic''
  dissipation},}\ }\href@noop {} {\bibfield  {journal} {\bibinfo  {journal}
  {JETP Lett.}\ }\textbf {\bibinfo {volume} {45}},\ \bibinfo {pages} {434}
  (\bibinfo {year} {1987})}\BibitemShut {NoStop}%
\bibitem [{\citenamefont {Little}(1967)}]{Little:1967}%
  \BibitemOpen
  \bibfield  {author} {\bibinfo {author} {\bibfnamefont {W.~A.}\ \bibnamefont
  {Little}},\ }\bibfield  {title} {\enquote {\bibinfo {title} {Decay of
  persistent currents in small superconductors},}\ }\href {\doibase
  10.1103/PhysRev.156.396} {\bibfield  {journal} {\bibinfo  {journal} {Phys.\
  Rev.}\ }\textbf {\bibinfo {volume} {156}},\ \bibinfo {pages} {396--403}
  (\bibinfo {year} {1967})}\BibitemShut {NoStop}%
\bibitem [{\citenamefont {Sahu}\ \emph {et~al.}(2009)\citenamefont {Sahu},
  \citenamefont {Bae}, \citenamefont {Rogachev}, \citenamefont {Pekker},
  \citenamefont {Wei}, \citenamefont {Shah}, \citenamefont {Goldbart},\ and\
  \citenamefont {Bezryadin}}]{Sahu:2009}%
  \BibitemOpen
  \bibfield  {author} {\bibinfo {author} {\bibfnamefont {M.}~\bibnamefont
  {Sahu}}, \bibinfo {author} {\bibfnamefont {M.-H.}\ \bibnamefont {Bae}},
  \bibinfo {author} {\bibfnamefont {A.}~\bibnamefont {Rogachev}}, \bibinfo
  {author} {\bibfnamefont {D.}~\bibnamefont {Pekker}}, \bibinfo {author}
  {\bibfnamefont {T.-C.}\ \bibnamefont {Wei}}, \bibinfo {author} {\bibfnamefont
  {N.}~\bibnamefont {Shah}}, \bibinfo {author} {\bibfnamefont {P.~M.}\
  \bibnamefont {Goldbart}}, \ and\ \bibinfo {author} {\bibfnamefont
  {A.}~\bibnamefont {Bezryadin}},\ }\bibfield  {title} {\enquote {\bibinfo
  {title} {Individual topological tunnelling events of a quantum field probed
  through their macroscopic consequences},}\ }\href {\doibase
  10.1038/NPHYS1276} {\bibfield  {journal} {\bibinfo  {journal} {Nat.\ Phys.}\
  }\textbf {\bibinfo {volume} {5}},\ \bibinfo {pages} {503--508} (\bibinfo
  {year} {2009})}\BibitemShut {NoStop}%
\bibitem [{\citenamefont {Li}\ \emph {et~al.}(2011)\citenamefont {Li},
  \citenamefont {Wu}, \citenamefont {Bomze}, \citenamefont {Borzenets},
  \citenamefont {Finkelstein},\ and\ \citenamefont {Chang}}]{Li:2011}%
  \BibitemOpen
  \bibfield  {author} {\bibinfo {author} {\bibfnamefont {P.}~\bibnamefont
  {Li}}, \bibinfo {author} {\bibfnamefont {P.~M.}\ \bibnamefont {Wu}}, \bibinfo
  {author} {\bibfnamefont {Yu.}\ \bibnamefont {Bomze}}, \bibinfo {author}
  {\bibfnamefont {I.~V.}\ \bibnamefont {Borzenets}}, \bibinfo {author}
  {\bibfnamefont {G.}~\bibnamefont {Finkelstein}}, \ and\ \bibinfo {author}
  {\bibfnamefont {A.~M.}\ \bibnamefont {Chang}},\ }\bibfield  {title} {\enquote
  {\bibinfo {title} {Switching currents limited by single phase slips in
  one-dimensional superconducting al nanowires},}\ }\href {\doibase
  10.1103/PhysRevLett.107.137004} {\bibfield  {journal} {\bibinfo  {journal}
  {Phys.\ Rev.\ Lett.}\ }\textbf {\bibinfo {volume} {107}},\ \bibinfo {pages}
  {137004} (\bibinfo {year} {2011})}\BibitemShut {NoStop}%
\bibitem [{\citenamefont {Aref}\ \emph {et~al.}(2012)\citenamefont {Aref},
  \citenamefont {Levchenko}, \citenamefont {Vakaryuk},\ and\ \citenamefont
  {Bezryadin}}]{Aref:2012}%
  \BibitemOpen
  \bibfield  {author} {\bibinfo {author} {\bibfnamefont {T.}~\bibnamefont
  {Aref}}, \bibinfo {author} {\bibfnamefont {A.}~\bibnamefont {Levchenko}},
  \bibinfo {author} {\bibfnamefont {V.}~\bibnamefont {Vakaryuk}}, \ and\
  \bibinfo {author} {\bibfnamefont {A.}~\bibnamefont {Bezryadin}},\ }\bibfield
  {title} {\enquote {\bibinfo {title} {Quantitative analysis of quantum phase
  slips in superconducting {Mo}$_{76}${Ge}$_{24}$ nanowires revealed by
  switching-current statistics},}\ }\href {\doibase 10.1103/PhysRevB.86.024507}
  {\bibfield  {journal} {\bibinfo  {journal} {Phys.\ Rev.~B}\ }\textbf
  {\bibinfo {volume} {86}},\ \bibinfo {pages} {024507} (\bibinfo {year}
  {2012})}\BibitemShut {NoStop}%
\bibitem [{\citenamefont {Bezryadin}(2012)}]{Bezryadin:2012}%
  \BibitemOpen
  \bibfield  {author} {\bibinfo {author} {\bibfnamefont {A.}~\bibnamefont
  {Bezryadin}},\ }\href@noop {} {\emph {\bibinfo {title} {Superconductivity in
  Nanowires: Fabrication and Quantum Transport}}}\ (\bibinfo  {publisher}
  {Wiley-VCH},\ \bibinfo {address} {69469 Weinheim, Germany},\ \bibinfo {year}
  {2012})\BibitemShut {NoStop}%
\bibitem [{\citenamefont {Giordano}(1988)}]{Giordano:1988}%
  \BibitemOpen
  \bibfield  {author} {\bibinfo {author} {\bibfnamefont {N.}~\bibnamefont
  {Giordano}},\ }\bibfield  {title} {\enquote {\bibinfo {title} {Evidence for
  macroscopic quantum tunneling in one-dimensional superconductors},}\ }\href
  {\doibase 10.1103/PhysRevLett.61.2137} {\bibfield  {journal} {\bibinfo
  {journal} {Phys.\ Rev.\ Lett.}\ }\textbf {\bibinfo {volume} {61}},\ \bibinfo
  {pages} {2137--2140} (\bibinfo {year} {1988})}\BibitemShut {NoStop}%
\bibitem [{\citenamefont {Bezryadin}\ \emph {et~al.}(2000)\citenamefont
  {Bezryadin}, \citenamefont {Lau},\ and\ \citenamefont
  {Tinkham}}]{Bezryadin:2000}%
  \BibitemOpen
  \bibfield  {author} {\bibinfo {author} {\bibfnamefont {A.}~\bibnamefont
  {Bezryadin}}, \bibinfo {author} {\bibfnamefont {C.~N.}\ \bibnamefont {Lau}},
  \ and\ \bibinfo {author} {\bibfnamefont {M.}~\bibnamefont {Tinkham}},\
  }\bibfield  {title} {\enquote {\bibinfo {title} {Quantum suppression of
  superconductivity in ultrathin nanowires},}\ }\href {\doibase
  10.1038/35010060} {\bibfield  {journal} {\bibinfo  {journal} {Nature
  (London)}\ }\textbf {\bibinfo {volume} {404}},\ \bibinfo {pages} {971--974}
  (\bibinfo {year} {2000})}\BibitemShut {NoStop}%
\bibitem [{\citenamefont {Lau}\ \emph {et~al.}(2001)\citenamefont {Lau},
  \citenamefont {Markovic}, \citenamefont {Bockrath}, \citenamefont
  {Bezryadin},\ and\ \citenamefont {Tinkham}}]{Lau:2001}%
  \BibitemOpen
  \bibfield  {author} {\bibinfo {author} {\bibfnamefont {C.~N.}\ \bibnamefont
  {Lau}}, \bibinfo {author} {\bibfnamefont {N.}~\bibnamefont {Markovic}},
  \bibinfo {author} {\bibfnamefont {M.}~\bibnamefont {Bockrath}}, \bibinfo
  {author} {\bibfnamefont {A.}~\bibnamefont {Bezryadin}}, \ and\ \bibinfo
  {author} {\bibfnamefont {M.}~\bibnamefont {Tinkham}},\ }\bibfield  {title}
  {\enquote {\bibinfo {title} {Quantum phase slips in superconducting
  nanowires},}\ }\href {\doibase 10.1103/PhysRevLett.87.217003} {\bibfield
  {journal} {\bibinfo  {journal} {Phys.\ Rev.\ Lett.}\ }\textbf {\bibinfo
  {volume} {87}},\ \bibinfo {pages} {217003} (\bibinfo {year}
  {2001})}\BibitemShut {NoStop}%
\bibitem [{\citenamefont {Arutyunov}\ \emph {et~al.}(2008)\citenamefont
  {Arutyunov}, \citenamefont {Golubev},\ and\ \citenamefont
  {Zaikin}}]{Arutyunov:2008}%
  \BibitemOpen
  \bibfield  {author} {\bibinfo {author} {\bibfnamefont {K.~Yu.}\ \bibnamefont
  {Arutyunov}}, \bibinfo {author} {\bibfnamefont {D.~S.}\ \bibnamefont
  {Golubev}}, \ and\ \bibinfo {author} {\bibfnamefont {A.~D.}\ \bibnamefont
  {Zaikin}},\ }\bibfield  {title} {\enquote {\bibinfo {title}
  {Superconductivity in one dimension},}\ }\href {\doibase
  10.1016/j.physrep.2008.04.009} {\bibfield  {journal} {\bibinfo  {journal}
  {Phys.\ Rep.}\ }\textbf {\bibinfo {volume} {464}},\ \bibinfo {pages} {1--70}
  (\bibinfo {year} {2008})}\BibitemShut {NoStop}%
\bibitem [{\citenamefont {Astafiev}\ \emph {et~al.}(2012)\citenamefont
  {Astafiev}, \citenamefont {Ioffe}, \citenamefont {Kafanov}, \citenamefont
  {Pashkin}, \citenamefont {Arutyunov}, \citenamefont {Shahar}, \citenamefont
  {Cohen},\ and\ \citenamefont {Tsai}}]{Astafiev:2012}%
  \BibitemOpen
  \bibfield  {author} {\bibinfo {author} {\bibfnamefont {O.~V.}\ \bibnamefont
  {Astafiev}}, \bibinfo {author} {\bibfnamefont {L.~B.}\ \bibnamefont {Ioffe}},
  \bibinfo {author} {\bibfnamefont {S.}~\bibnamefont {Kafanov}}, \bibinfo
  {author} {\bibfnamefont {Yu.~A.}\ \bibnamefont {Pashkin}}, \bibinfo {author}
  {\bibfnamefont {K.~Yu.}\ \bibnamefont {Arutyunov}}, \bibinfo {author}
  {\bibfnamefont {D.}~\bibnamefont {Shahar}}, \bibinfo {author} {\bibfnamefont
  {O.}~\bibnamefont {Cohen}}, \ and\ \bibinfo {author} {\bibfnamefont {J.~S.}\
  \bibnamefont {Tsai}},\ }\bibfield  {title} {\enquote {\bibinfo {title}
  {Coherent quantum phase slip},}\ }\href {\doibase 10.1038/nature10930}
  {\bibfield  {journal} {\bibinfo  {journal} {Nature (London)}\ }\textbf
  {\bibinfo {volume} {484}},\ \bibinfo {pages} {355--358} (\bibinfo {year}
  {2012})}\BibitemShut {NoStop}%
\bibitem [{\citenamefont {Peltonen}\ \emph {et~al.}(2013)\citenamefont
  {Peltonen}, \citenamefont {Astafiev}, \citenamefont {Korneeva}, \citenamefont
  {Voronov}, \citenamefont {Korneev}, \citenamefont {Charaev}, \citenamefont
  {Semenov}, \citenamefont {Golt'sman}, \citenamefont {Ioffe}, \citenamefont
  {Klapwijk},\ and\ \citenamefont {Tsai}}]{Peltonen:2013}%
  \BibitemOpen
  \bibfield  {author} {\bibinfo {author} {\bibfnamefont {J.~T.}\ \bibnamefont
  {Peltonen}}, \bibinfo {author} {\bibfnamefont {O.~V.}\ \bibnamefont
  {Astafiev}}, \bibinfo {author} {\bibfnamefont {Yu.~P.}\ \bibnamefont
  {Korneeva}}, \bibinfo {author} {\bibfnamefont {B.~M.}\ \bibnamefont
  {Voronov}}, \bibinfo {author} {\bibfnamefont {A.~A.}\ \bibnamefont
  {Korneev}}, \bibinfo {author} {\bibfnamefont {I.~M.}\ \bibnamefont
  {Charaev}}, \bibinfo {author} {\bibfnamefont {A.~V.}\ \bibnamefont
  {Semenov}}, \bibinfo {author} {\bibfnamefont {G.~N.}\ \bibnamefont
  {Golt'sman}}, \bibinfo {author} {\bibfnamefont {L.~B.}\ \bibnamefont
  {Ioffe}}, \bibinfo {author} {\bibfnamefont {T.~M.}\ \bibnamefont {Klapwijk}},
  \ and\ \bibinfo {author} {\bibfnamefont {J.~S.}\ \bibnamefont {Tsai}},\
  }\bibfield  {title} {\enquote {\bibinfo {title} {Coherent flux tunneling
  through nbn nanowires},}\ }\href {\doibase 10.1103/PhysRevB.88.220506}
  {\bibfield  {journal} {\bibinfo  {journal} {Phys.\ Rev.~B}\ }\textbf
  {\bibinfo {volume} {88}},\ \bibinfo {pages} {220506} (\bibinfo {year}
  {2013})}\BibitemShut {NoStop}%
\bibitem [{\citenamefont {Belkin}\ \emph {et~al.}(2011)\citenamefont {Belkin},
  \citenamefont {Brenner}, \citenamefont {Aref}, \citenamefont {Ku},\ and\
  \citenamefont {Bezryadin}}]{Belkin:2011}%
  \BibitemOpen
  \bibfield  {author} {\bibinfo {author} {\bibfnamefont {A.}~\bibnamefont
  {Belkin}}, \bibinfo {author} {\bibfnamefont {M.}~\bibnamefont {Brenner}},
  \bibinfo {author} {\bibfnamefont {T.}~\bibnamefont {Aref}}, \bibinfo {author}
  {\bibfnamefont {J.}~\bibnamefont {Ku}}, \ and\ \bibinfo {author}
  {\bibfnamefont {A.}~\bibnamefont {Bezryadin}},\ }\bibfield  {title} {\enquote
  {\bibinfo {title} {Little-parks oscillations at low temperatures: Gigahertz
  resonator method},}\ }\href {\doibase 10.1063/1.3593482} {\bibfield
  {journal} {\bibinfo  {journal} {Appl.\ Phys.\ Lett.}\ }\textbf {\bibinfo
  {volume} {98}},\ \bibinfo {pages} {242504} (\bibinfo {year}
  {2011})}\BibitemShut {NoStop}%
\bibitem [{\citenamefont {Little}\ and\ \citenamefont
  {Parks}(1962)}]{Little:1962}%
  \BibitemOpen
  \bibfield  {author} {\bibinfo {author} {\bibfnamefont {W.~A.}\ \bibnamefont
  {Little}}\ and\ \bibinfo {author} {\bibfnamefont {R.~D.}\ \bibnamefont
  {Parks}},\ }\bibfield  {title} {\enquote {\bibinfo {title} {Observation of
  quantum periodicity in the transition temperature of a superconducting
  cylinder},}\ }\href {\doibase 10.1103/PhysRevLett.9.9} {\bibfield  {journal}
  {\bibinfo  {journal} {Phys.\ Rev.\ Lett.}\ }\textbf {\bibinfo {volume} {9}},\
  \bibinfo {pages} {9--12} (\bibinfo {year} {1962})}\BibitemShut {NoStop}%
\bibitem [{\citenamefont {Parks}\ and\ \citenamefont
  {Little}(1964)}]{Parks:1964}%
  \BibitemOpen
  \bibfield  {author} {\bibinfo {author} {\bibfnamefont {R.~D.}\ \bibnamefont
  {Parks}}\ and\ \bibinfo {author} {\bibfnamefont {W.~A.}\ \bibnamefont
  {Little}},\ }\bibfield  {title} {\enquote {\bibinfo {title} {Fluxoid
  quantization in a multiply-connected superconductor},}\ }\href {\doibase
  10.1103/PhysRev.133.A97} {\bibfield  {journal} {\bibinfo  {journal} {Phys.\
  Rev.}\ }\textbf {\bibinfo {volume} {133}},\ \bibinfo {pages} {A97--A103}
  (\bibinfo {year} {1964})}\BibitemShut {NoStop}%
\bibitem [{\citenamefont {Hopkins}\ \emph {et~al.}(2005)\citenamefont
  {Hopkins}, \citenamefont {Pekker}, \citenamefont {Goldbart},\ and\
  \citenamefont {Bezryadin}}]{Hopkins:2005}%
  \BibitemOpen
  \bibfield  {author} {\bibinfo {author} {\bibfnamefont {D.~S.}\ \bibnamefont
  {Hopkins}}, \bibinfo {author} {\bibfnamefont {D.}~\bibnamefont {Pekker}},
  \bibinfo {author} {\bibfnamefont {P.~M.}\ \bibnamefont {Goldbart}}, \ and\
  \bibinfo {author} {\bibfnamefont {A.}~\bibnamefont {Bezryadin}},\ }\bibfield
  {title} {\enquote {\bibinfo {title} {Quantum interference device made by dna
  templating of superconducting nanowires},}\ }\href {\doibase
  10.1126/science.1111307} {\bibfield  {journal} {\bibinfo  {journal}
  {Science}\ }\textbf {\bibinfo {volume} {308}},\ \bibinfo {pages} {1762--1765}
  (\bibinfo {year} {2005})}\BibitemShut {NoStop}%
\bibitem [{\citenamefont {Kurkij\"arvi}(1972)}]{Kurkij:1972}%
  \BibitemOpen
  \bibfield  {author} {\bibinfo {author} {\bibfnamefont {J.}~\bibnamefont
  {Kurkij\"arvi}},\ }\bibfield  {title} {\enquote {\bibinfo {title} {Intrinsic
  fluctuations in a superconducting ring closed with a josephson junction},}\
  }\href {\doibase 10.1103/PhysRevB.6.832} {\bibfield  {journal} {\bibinfo
  {journal} {Phys.\ Rev.~B}\ }\textbf {\bibinfo {volume} {6}},\ \bibinfo
  {pages} {832--835} (\bibinfo {year} {1972})}\BibitemShut {NoStop}%
\bibitem [{\citenamefont {Garg}(1995)}]{Garg:1995}%
  \BibitemOpen
  \bibfield  {author} {\bibinfo {author} {\bibfnamefont {A.}~\bibnamefont
  {Garg}},\ }\bibfield  {title} {\enquote {\bibinfo {title} {Escape-field
  distribution for escape from a metastable potential well subject to a
  steadily increasing bias field},}\ }\href {\doibase
  10.1103/PhysRevB.51.15592} {\bibfield  {journal} {\bibinfo  {journal} {Phys.\
  Rev.~B}\ }\textbf {\bibinfo {volume} {51}},\ \bibinfo {pages} {15592--15595}
  (\bibinfo {year} {1995})}\BibitemShut {NoStop}%
\bibitem [{\citenamefont {Fulton}\ and\ \citenamefont
  {Dunkleberger}(1974)}]{Fulton:1974}%
  \BibitemOpen
  \bibfield  {author} {\bibinfo {author} {\bibfnamefont {T.~A.}\ \bibnamefont
  {Fulton}}\ and\ \bibinfo {author} {\bibfnamefont {L.~N.}\ \bibnamefont
  {Dunkleberger}},\ }\bibfield  {title} {\enquote {\bibinfo {title} {Lifetime
  of the zero-voltage state in josephson tunnel junctions},}\ }\href {\doibase
  10.1103/PhysRevB.9.4760} {\bibfield  {journal} {\bibinfo  {journal} {Phys.\
  Rev.~B}\ }\textbf {\bibinfo {volume} {9}},\ \bibinfo {pages} {4760--4768}
  (\bibinfo {year} {1974})}\BibitemShut {NoStop}%
\bibitem [{\citenamefont {Anderson}\ and\ \citenamefont
  {Dayem}(1964)}]{Anderson:1964}%
  \BibitemOpen
  \bibfield  {author} {\bibinfo {author} {\bibfnamefont {P.~W.}\ \bibnamefont
  {Anderson}}\ and\ \bibinfo {author} {\bibfnamefont {A.~H.}\ \bibnamefont
  {Dayem}},\ }\bibfield  {title} {\enquote {\bibinfo {title} {Radio-frequency
  effects in superconducting thin film bridges},}\ }\href {\doibase
  10.1103/PhysRevLett.13.195} {\bibfield  {journal} {\bibinfo  {journal}
  {Phys.\ Rev.\ Lett.}\ }\textbf {\bibinfo {volume} {13}},\ \bibinfo {pages}
  {195--197} (\bibinfo {year} {1964})}\BibitemShut {NoStop}%
\bibitem [{\citenamefont {Shah}\ \emph {et~al.}(2008)\citenamefont {Shah},
  \citenamefont {Pekker},\ and\ \citenamefont {Goldbart}}]{Shah:2008}%
  \BibitemOpen
  \bibfield  {author} {\bibinfo {author} {\bibfnamefont {N.}~\bibnamefont
  {Shah}}, \bibinfo {author} {\bibfnamefont {D.}~\bibnamefont {Pekker}}, \ and\
  \bibinfo {author} {\bibfnamefont {P.~M.}\ \bibnamefont {Goldbart}},\
  }\bibfield  {title} {\enquote {\bibinfo {title} {Inherent stochasticity of
  superconductor-resistor switching behavior in nanowires},}\ }\href {\doibase
  10.1103/PhysRevLett.101.207001} {\bibfield  {journal} {\bibinfo  {journal}
  {Phys.\ Rev.\ Lett.}\ }\textbf {\bibinfo {volume} {101}},\ \bibinfo {pages}
  {207001} (\bibinfo {year} {2008})}\BibitemShut {NoStop}%
\bibitem [{\citenamefont {Pekker}\ \emph {et~al.}(2009)\citenamefont {Pekker},
  \citenamefont {Shah}, \citenamefont {Sahu}, \citenamefont {Bezryadin},\ and\
  \citenamefont {Goldbart}}]{Pekker:2009}%
  \BibitemOpen
  \bibfield  {author} {\bibinfo {author} {\bibfnamefont {D.}~\bibnamefont
  {Pekker}}, \bibinfo {author} {\bibfnamefont {N.}~\bibnamefont {Shah}},
  \bibinfo {author} {\bibfnamefont {M.}~\bibnamefont {Sahu}}, \bibinfo {author}
  {\bibfnamefont {A.}~\bibnamefont {Bezryadin}}, \ and\ \bibinfo {author}
  {\bibfnamefont {P.~M.}\ \bibnamefont {Goldbart}},\ }\bibfield  {title}
  {\enquote {\bibinfo {title} {Stochastic dynamics of phase-slip trains and
  superconductive-resistive switching in current-biased nanowires},}\ }\href
  {\doibase 10.1103/PhysRevB.80.214525} {\bibfield  {journal} {\bibinfo
  {journal} {Phys.\ Rev.~B}\ }\textbf {\bibinfo {volume} {80}},\ \bibinfo
  {pages} {214525} (\bibinfo {year} {2009})}\BibitemShut {NoStop}%
\bibitem [{\citenamefont {Sch\"{o}n}\ and\ \citenamefont
  {Zaikin}(1990)}]{Zaikin:1990}%
  \BibitemOpen
  \bibfield  {author} {\bibinfo {author} {\bibfnamefont {G.}~\bibnamefont
  {Sch\"{o}n}}\ and\ \bibinfo {author} {\bibfnamefont {A.~D.}\ \bibnamefont
  {Zaikin}},\ }\bibfield  {title} {\enquote {\bibinfo {title} {Quantum coherent
  effects, phase transitions, and the dissipative dynamics of ultra small
  tunnel junctions},}\ }\href@noop {} {\bibfield  {journal} {\bibinfo
  {journal} {Phys.\ Rep.}\ }\textbf {\bibinfo {volume} {198}},\ \bibinfo
  {pages} {237--412} (\bibinfo {year} {1990})}\BibitemShut {NoStop}%
\bibitem [{\citenamefont {Brenner}\ \emph {et~al.}(2012)\citenamefont
  {Brenner}, \citenamefont {Roy}, \citenamefont {Shah},\ and\ \citenamefont
  {Bezryadin}}]{Brenner:2012}%
  \BibitemOpen
  \bibfield  {author} {\bibinfo {author} {\bibfnamefont {M.~W.}\ \bibnamefont
  {Brenner}}, \bibinfo {author} {\bibfnamefont {D.}~\bibnamefont {Roy}},
  \bibinfo {author} {\bibfnamefont {N.}~\bibnamefont {Shah}}, \ and\ \bibinfo
  {author} {\bibfnamefont {A.}~\bibnamefont {Bezryadin}},\ }\bibfield  {title}
  {\enquote {\bibinfo {title} {Dynamics of superconducting nanowires shunted
  with an external resistor},}\ }\href {\doibase 10.1103/PhysRevB.85.224507}
  {\bibfield  {journal} {\bibinfo  {journal} {Phys.\ Rev.~B}\ }\textbf
  {\bibinfo {volume} {85}},\ \bibinfo {pages} {224507} (\bibinfo {year}
  {2012})}\BibitemShut {NoStop}%
\bibitem [{\citenamefont {Likharev}(1979)}]{Likharev:1979}%
  \BibitemOpen
  \bibfield  {author} {\bibinfo {author} {\bibfnamefont {K.}~\bibnamefont
  {Likharev}},\ }\bibfield  {title} {\enquote {\bibinfo {title}
  {Superconducting weak links},}\ }\href {\doibase 10.1103/RevModPhys.51.101}
  {\bibfield  {journal} {\bibinfo  {journal} {Rev.\ Mod.\ Phys.}\ }\textbf
  {\bibinfo {volume} {51}},\ \bibinfo {pages} {101--159} (\bibinfo {year}
  {1979})}\BibitemShut {NoStop}%
\bibitem [{\citenamefont {Rogachev}\ \emph {et~al.}(2005)\citenamefont
  {Rogachev}, \citenamefont {Bollinger},\ and\ \citenamefont
  {Bezryadin}}]{Rogachev:2005}%
  \BibitemOpen
  \bibfield  {author} {\bibinfo {author} {\bibfnamefont {A.}~\bibnamefont
  {Rogachev}}, \bibinfo {author} {\bibfnamefont {A.~T.}\ \bibnamefont
  {Bollinger}}, \ and\ \bibinfo {author} {\bibfnamefont {A.}~\bibnamefont
  {Bezryadin}},\ }\bibfield  {title} {\enquote {\bibinfo {title} {Influence of
  high magnetic fields on the superconducting transition of one-dimensional
  {N}b and {M}o{G}e nanowires},}\ }\href {\doibase
  10.1103/PhysRevLett.94.017004} {\bibfield  {journal} {\bibinfo  {journal}
  {Phys.\ Rev.\ Lett.}\ }\textbf {\bibinfo {volume} {94}},\ \bibinfo {pages}
  {017004} (\bibinfo {year} {2005})}\BibitemShut {NoStop}%
\bibitem [{\citenamefont {White}\ \emph {et~al.}(1993)\citenamefont {White},
  \citenamefont {Kapitulnik},\ and\ \citenamefont {Beasley}}]{White:1993}%
  \BibitemOpen
  \bibfield  {author} {\bibinfo {author} {\bibfnamefont {W.~R.}\ \bibnamefont
  {White}}, \bibinfo {author} {\bibfnamefont {A.}~\bibnamefont {Kapitulnik}}, \
  and\ \bibinfo {author} {\bibfnamefont {M.~R.}\ \bibnamefont {Beasley}},\
  }\bibfield  {title} {\enquote {\bibinfo {title} {Collective vortex motion in
  a-{M}o{G}e superconducting thin films},}\ }\href {\doibase
  10.1103/PhysRevLett.70.670} {\bibfield  {journal} {\bibinfo  {journal}
  {Phys.\ Rev.\ Lett.}\ }\textbf {\bibinfo {volume} {70}},\ \bibinfo {pages}
  {670--673} (\bibinfo {year} {1993})}\BibitemShut {NoStop}%
\bibitem [{\citenamefont {Caldeira}\ and\ \citenamefont
  {Leggett}(1981)}]{Caldeira:1981}%
  \BibitemOpen
  \bibfield  {author} {\bibinfo {author} {\bibfnamefont {A.~O.}\ \bibnamefont
  {Caldeira}}\ and\ \bibinfo {author} {\bibfnamefont {A.~J.}\ \bibnamefont
  {Leggett}},\ }\bibfield  {title} {\enquote {\bibinfo {title} {Influence of
  dissipation on quantum tunneling in macroscopic systems},}\ }\href {\doibase
  10.1103/PhysRevLett.46.211} {\bibfield  {journal} {\bibinfo  {journal}
  {Phys.\ Rev.\ Lett.}\ }\textbf {\bibinfo {volume} {46}},\ \bibinfo {pages}
  {211--214} (\bibinfo {year} {1981})}\BibitemShut {NoStop}%
\bibitem [{\citenamefont {Nazarov}(1999)}]{Nazarov:1999}%
  \BibitemOpen
  \bibfield  {author} {\bibinfo {author} {\bibfnamefont {Yu.~V.}\ \bibnamefont
  {Nazarov}},\ }\bibfield  {title} {\enquote {\bibinfo {title} {Coulomb
  blockade without tunnel junctions},}\ }\href {\doibase
  10.1103/PhysRevLett.82.1245} {\bibfield  {journal} {\bibinfo  {journal}
  {Phys.\ Rev.\ Lett.}\ }\textbf {\bibinfo {volume} {82}},\ \bibinfo {pages}
  {1245--1248} (\bibinfo {year} {1999})}\BibitemShut {NoStop}%
\bibitem [{\citenamefont {Golubev}\ and\ \citenamefont
  {Zaikin}(2001)}]{Golubev:2001}%
  \BibitemOpen
  \bibfield  {author} {\bibinfo {author} {\bibfnamefont {D.~S.}\ \bibnamefont
  {Golubev}}\ and\ \bibinfo {author} {\bibfnamefont {A.~D.}\ \bibnamefont
  {Zaikin}},\ }\bibfield  {title} {\enquote {\bibinfo {title} {Coulomb
  interaction and quantum transport through a coherent scatterer},}\ }\href
  {\doibase 10.1103/PhysRevLett.86.4887} {\bibfield  {journal} {\bibinfo
  {journal} {Phys.\ Rev.\ Lett.}\ }\textbf {\bibinfo {volume} {86}},\ \bibinfo
  {pages} {4887--4890} (\bibinfo {year} {2001})}\BibitemShut {NoStop}%
\bibitem [{\citenamefont {Bollinger}\ \emph {et~al.}(2006)\citenamefont
  {Bollinger}, \citenamefont {Rogachev},\ and\ \citenamefont
  {Bezryadin}}]{Bollinger:2006}%
  \BibitemOpen
  \bibfield  {author} {\bibinfo {author} {\bibfnamefont {A.~T.}\ \bibnamefont
  {Bollinger}}, \bibinfo {author} {\bibfnamefont {A.}~\bibnamefont {Rogachev}},
  \ and\ \bibinfo {author} {\bibfnamefont {A}~\bibnamefont {Bezryadin}},\
  }\bibfield  {title} {\enquote {\bibinfo {title} {Dichotomy in short
  superconducting nanowires: Thermal phase slippage vs. coulomb blockade},}\
  }\href {\doibase 10.1209/epl/i2006-10275-5} {\bibfield  {journal} {\bibinfo
  {journal} {Europhys.\ Lett.}\ }\textbf {\bibinfo {volume} {76}},\ \bibinfo
  {pages} {505} (\bibinfo {year} {2006})}\BibitemShut {NoStop}%
\end{thebibliography}%

\end{document}